\documentclass[aps, prd, 10pt, notitlepage, superscriptaddress, nofootinbib,numbers]{revtex4-1}

\usepackage[utf8]{inputenc}
\usepackage[T1]{fontenc}
\usepackage{anyfontsize}

\usepackage{amsmath}
\usepackage{amssymb}
\usepackage{amsfonts}
\usepackage{mathrsfs}

\usepackage{microtype}

\usepackage{graphicx}
\usepackage{epsfig}

\usepackage[dvipsnames]{xcolor}
\usepackage{hyperref}
\hypersetup{
    colorlinks=true,
    citecolor=Purple,
    linkcolor=Purple,
    urlcolor=Purple,
    linktocpage=true,
    breaklinks=true
}

\def\lal{&&\ {}}

\def\beq{\begin{equation}}
\def\eeq{\end{equation}}
\def\bear{\begin{eqnarray}}
\def\bearr{\begin{eqnarray} \lal}
\def\ear{\end{eqnarray}}
\def\earn{\nonumber \end{eqnarray}}

\usepackage{orcidlink}

\usepackage[capitalize]{cleveref}

\begin{document}

\title{Generalized black-bounces solutions in $f(R)$ gravity and their field sources}
\author{Marcos V. de S. Silva\footnote{Author to whom any correspondence should be addressed.}}
\email{marcosvinicius@fisica.ufc.br}
\affiliation{Departamento de F\'isica, Universidade Federal do Cear\'a, Caixa Postal 6030, Campus do Pici, 60455-760 Fortaleza, Cear\'a, Brazil.}
\author{T. M. Crispim}
\email{tiago.crispim@fisica.ufc.br}
\affiliation{Departamento de F\'isica, Universidade Federal do Cear\'a, Caixa Postal 6030, Campus do Pici, 60455-760 Fortaleza, Cear\'a, Brazil.}

\author{G. Alencar}
\email{geova@fisica.ufc.br}
\affiliation{Departamento de F\'isica, Universidade Federal do Cear\'a, Caixa Postal 6030, Campus do Pici, 60455-760 Fortaleza, Cear\'a, Brazil.}
\author{R. R. Landim}
\email{renan@fisica.ufc.br}
\affiliation{Departamento de F\'isica, Universidade Federal do Cear\'a, Caixa Postal 6030, Campus do Pici, 60455-760 Fortaleza, Cear\'a, Brazil.}

\author{Manuel E. Rodrigues}
\email{esialg@gmail.com}
\affiliation{Faculdade de F\'{i}sica, Programa de P\'{o}s-Gradua\c{c}\~{a}o em F\'{i}sica, Universidade Federal do Par\'{a}, 66075-110, Bel\'{e}m, Par\'{a}, Brazill}
\affiliation{Faculdade de Ci\^{e}ncias Exatas e Tecnologia, Universidade Federal do Par\'{a}, Campus Universit\'{a}rio de Abaetetuba, 68440-000, Abaetetuba, Par\'{a}, Brazil}



\date{\today}

\begin{abstract}
In this work, following our recent findings in \cite{Alencar:2024nxi}, we extend our analysis to explore the generalization of spherically symmetric and static black-bounce solutions, known from General Relativity, within the framework of the $f(R)$ theory in the metric formalism. We develop a general approach to determine the sources for any model where $f(R) = R + H(R)$, provided that the corresponding source for the bounce metric in General Relativity is known. As a result, we demonstrate that black-bounce solutions can emerge from this theory when considering the coupling of $f(R)$ gravity with nonlinear electrodynamics and a partially phantom scalar field. We also analyzed the energy conditions of these solutions and found that, unlike in General Relativity, it is possible to satisfy all energy conditions in certain regions of space-time.
\end{abstract}


\maketitle

\section{Introduction}
General Relativity (GR) is the most widely accepted theory of gravitation. This theory describes astrophysical phenomena with great precision. One of the most impressive predictions of GR is the existence of black holes (BHs). These objects stand out due to their causal structure and have significant relevance in the current observational scenario, as we can now access astrophysical information about them through gravitational wave observations and BH images \cite{LIGOScientific:2016aoc,EventHorizonTelescope:2019dse}. Due to their intense gravitational field, BHs are the most suitable laboratories to detect the need for modifications in existing gravitational theories. Despite their great relevance, BHs bring with them one of the biggest problems of classical gravitation: the presence of singularities. These singularities are points, or sets of points, in space-time where geodesics are interrupted, making it difficult to describe the physics in these regions \cite{Bronnikov:2012wsj}.

The so-called black-bounce (BB) space-times are solutions of GR that aim to address the presence of singularities within the event horizon of BHs. To tackle this, Simpson and Visser (SV) \cite{Simpson:2018tsi} proposed a metric that represents the minimal modification of the Schwarzschild metric capable of producing a regular space-time. Specifically, they replaced the radial coordinate $r$ for $r \to \sqrt{r^2 + a^2}$. This resulted in a geometry that, depending on the value of the parameter $a$, interpolates between a regular BH and a traversable wormhole, thereby attempting to unify these two seemingly distinct space-times into a single solution.

Such solutions have been gaining increasing interest, with this procedure being applied in a variety of contexts \cite{Lobo:2020ffi,Rodrigues:2022mdm,Rodrigues:2022rfj,Rodrigues:2025plw}, including cylindrical geometries \cite{Lima:2022pvc,Lima:2023arg,Lima:2023jtl,Bronnikov:2023aya}, braneworld scenarios \cite{Crispim:2024nou,Crispim:2024yjz}, as well as their effects on gravitational lensing \cite{Islam:2021ful,Nascimento:2020ime,Ghosh:2022mka,Tsukamoto:2020bjm,Tsukamoto:2021caq}, gravitational wave echoes \cite{Yang:2021cvh}, quasinormal modes \cite{Franzin:2022iai}, and BH shadows \cite{Guerrero:2021ues,Guo:2021wid,Olmo:2023lil,Lima:2021las}.

On the other hand, it is well known that, in the context of GR, regular BHs \cite{bardeen1968,dymnikova1992,hayward2006} and traversable wormholes \cite{ellis1973, bronnikov1973,morris1988} are characterized by being supported by non-conventional field sources \cite{Rodrigues:2018bdc,Ayon-Beato:2000mjt,Bronnikov:2018vbs,Crispim:2024dgd,deSSilva:2024gdc,Crispim:2024lzf}, that is, field sources that violate at least one of the energy conditions. Thus, since BB are geometries that transition between wormholes and regular BHs, it is natural that such solutions are also supported by sources that violate the energy conditions. Specifically in the context of GR, the search for field sources for BB metrics is currently a widely studied area.

Specifically, in \cite{Bronnikov:2021uta}, by combining a phantom scalar field, that is, one with a negative kinetic term, with a nonlinear electrodynamics theory, both minimally coupled with gravity, the first source for the SV geometry was found. From there, a series of studies on the topic were developed, such as in \cite{Alencar:2024yvh}, where an in-depth study on sources of BB in GR was developed. Specifically, the authors demonstrated that nonlinear electrodynamics, both electric and magnetic, combined with a partially phantom scalar field, can serve as a source for various BB geometries, both spherically symmetric and cylindrically symmetric. Similar approaches were developed in \cite{Alencar:2025jvl,Rodrigues:2023vtm, Pereira:2024rtv,Bronnikov:2022bud,Pereira:2024gsl,Junior:2025sjr}. In \cite{Alencar:2025jvl}, a procedure is presented to construct any generic spherically symmetric solution of this type of geometry using nonlinear electrodynamics, making it possible to construct the already known BB as well as modifications that represent more general solutions than those initially proposed by SV.

The knowledge about the source fields of these solutions is of great importance, as they modify the properties of these solutions. In \cite{Franzin:2023slm}, for example, the authors show how the quasinormal modes of regular solutions are altered when considering test fields and perturbations in the source fields. In \cite{Silva:2024fpn}, the authors investigate how photons propagate in the space-time of a BB when considering nonlinear electrodynamics. In \cite{dePaula:2024yzy}, it is shown that, in the presence of nonlinear electrodynamics, photons can behave as causal tachyons or acausal bradyons. Nonlinear electrodynamics also modifies the thermodynamics of BHs \cite{Rodrigues:2022qdp}. All these properties have only been well studied due to the knowledge about the source fields of these solutions.

In this context, this work aims to conduct a detailed study on field sources for BB geometry in $f(R)$ gravity, where we demonstrated that BB solutions can emerge from this theory when considering the coupling of $f(R)$ gravity with nonlinear electrodynamics and a partially phantom or canonical scalar field, depending on the choice of parameters. The $f(R)$ theory has been gaining increasing attention in the study of compact objects due to its corrections to GR, with numerous studies conducted on BHs \cite{Capozziello:2007id,Sebastiani:2010kv,Capozziello:2007wc, Kainulainen:2007bt,Nashed:2019tuk,Multamaki:2006zb,Rois:2024iiu,Nojiri:2024qgx,Elizalde:2020icc,Nojiri:2017kex,Nojiri:2014jqa}, wormholes \cite{DeBenedictis:2012qz,Samanta:2019tjb,Bhattacharya:2015oma,Bambi:2015zch}, regular BHs \cite{Rodrigues:2015ayd,Rodrigues:2016fym,Santos:2023vox,Tangphati:2023xnw,Rodrigues:2019xrc}, and more recently, BB \cite{Alencar:2024nxi,Fabris:2023opv,Junior:2023qaq,Junior:2024vrv,Junior:2024cbb,Rois:2024qzm}. However, the study of BB sources in $f(R)$ theories is a topic that has been very sparsely explored in the literature. In fact, our work is the first to explore this topic.

In this work, we adopt two distinct procedures: the first consists of proposing a specific form for the function $f(R)$ and the scalar field $\phi(r)$ and, from that, finding the form for the Lagrangian of the nonlinear electromagnetic field $L(F)$, the potential associated with the scalar field $V(\phi)$ and the function $h(\phi)$. The second, similar to the first, consists in proposing a form for the derivative of the function $f(R)$, $f_R$, such that $f(R)$ will depend on the space-time, in our case, the SV space-time. With the introduction of the corrections coming from the modification in the gravitational action, we conclude that contrary to GR, it is always possible to satisfy all energy conditions in some region of space-time.

This paper is organized as follows: In SEC. \ref{S:spacetime}, we will provide a brief review of the properties of SV space-time. The general equations and sources for the different cases analyzed here will be presented in SEC. \ref{S:Fields}, while the energy conditions will be discussed in SEC. \ref{S:energy}. Finally, in SEC. \ref{S:conclusions}, our conclusions and final analyses will be presented.

We adopt the metric signature $(+, -, -, -)$ and use the geometric units where $8\pi G = c = 1$.
\section{The Space-time}\label{S:spacetime}
Our goal is to verify whether BB can emerge as a solution to modified theories of gravity. To this end, we will consider the SV model \cite{Simpson:2018tsi}, which is described by a line element of the form:
\begin{equation}
    ds^2=A(r)dt^2-A(r)^{-1}dr^2-\Sigma(r)^2\left(d\theta^2+\sin^2\theta d\varphi^2\right),\label{line}
\end{equation}
where
\begin{equation}
    A(r)=1-\frac{2m}{\sqrt{r^2+a^2}},\qquad \mbox{and} \qquad \Sigma(r)^2=r^2+a^2,
\end{equation}
where $m$ is understood as the Schwarzschild mass and $a >0$ is a parameter responsible for making this geometry globally regular. Moreover, the parameter $a$ controls the interpolation of the geometry between a regular BH and a wormhole, such that
\begin{itemize}
    \item For $a < 2m$, we have a regular BH with horizons at $r_{\pm} = \pm \sqrt{4m^2 - a^2}$;
    \item For $a =2m$, we obtain what is known as a one-way wormhole, with a null throat located at  $r = 0$;
    \item For $a > 2m$, we have a traversable wormhole.
\end{itemize}

To infer the regularity of this geometry, we can analyze the curvature invariants. In this case, the Kretschmann and Ricci scalars are given, respectively, by
\begin{eqnarray}
    K &=& \frac{4}{(r^2 + a^2)^5}\left\{\sqrt{r^2 + a^2}[8a^2m(r^2 - a^2)] + 3a^4(r^2 + a^2) + 3m^2(3a^4- 4a^2r^2 + 4r^4)\right\},\label{Kre}\\
    R &=& \frac{2a^2(3m - \sqrt{r^2 + a^2})}{(r^2 + a^2)^{5/2}}.\label{RicciScalar}
\end{eqnarray}

At the origin of the radial coordinate, the curvature invariants take finite values given by
\begin{equation}
        \lim_{r \to 0} K = \frac{4(3a^2 + 9m^2 - 8am)}{a^6},\qquad
        \lim_{r \to 0} R = \frac{2(-3m + a)}{a^3},
\end{equation}
and are asymptotically zero since this space-time is asymptotically flat.

We can further analyze the regularity of space-time from another perspective. Thus, the geodesic equation for the $r$ coordinate, assuming, without loss of generality, $\theta = \pi/2$, is given by
\begin{equation}\label{geodesic}
    A(r)\dot{t}^2 - A^{-1}(r)\dot{r}^2 - \Sigma^2(r)\dot{\varphi}^2 = \epsilon,
\end{equation}
where the dot is a derivative with respect to affine parameter $\lambda$ and
\begin{eqnarray}
    \epsilon &=& 1 \;\; \text{for timelike geodesics},\\
    \epsilon &=& 0 \;\; \text{for null geodesics},\\
    \epsilon &=& -1 \;\; \text{for spacelike geodesics}.
\end{eqnarray}

The symmetries of the metric allow us to associate conserved quantities with the Killing vectors of the $t$ and $\phi$ coordinates, such that we have
\begin{eqnarray}
    \dot{t}A(r) &=& \text{constant} = E,\\
    \dot{\phi}\Sigma^2(r) &=& \text{constant} = l.
\end{eqnarray}
Substituting this into the equation \eqref{geodesic}
\begin{equation}
  \label{rgeo}  \dot{r}^2(\lambda) = E^2 - A(r)\left(\frac{l^2}{\Sigma^2(r)} - \epsilon\right).
\end{equation}
To determine where the geodesic is complete, that is, where it can be extended, we must solve the equation above. However, solving it analytically is an extremely challenging task. Instead, we will consider three important regimes: $r \to \infty$, $r \to r_h$, and $r \to 0$. For the first case, we have
\begin{equation}
    \dot{r}^2(\lambda) \sim E^2 + \epsilon -\frac{2m\epsilon}{r} + O\left(\frac{1}{r^2}\right) \sim E^2 + \epsilon.
\end{equation}
The solution is then
\begin{equation}
    r(\lambda) \sim \bar{c} \pm \lambda\sqrt{E^2 + \epsilon},
\end{equation}
where $\bar{c}$ is a constant of integration. Examining the equation above, we clearly see that for $\lambda \to \infty$ we have $r(\lambda) \to \infty$, so $r(\lambda)$ is extensible into the infinite future \cite{Rodrigues:2023fps}.

 For the second case, where $r$ tends to the value of the radius of the horizon, we have $A(r_h) \to 0$. Therefore, from the equation \eqref{rgeo} we have
\begin{eqnarray}
    r(\lambda) \sim \tilde{c} \pm |E|\lambda,
\end{eqnarray}
where it is trivially satisfied that the geodesics are extensible for this case.
 
 For the last case, where $r \to 0$, we have
\begin{equation}
    \dot{r}^2(\lambda) \sim c_0 + c_1r^2 + O(r^3),
\end{equation}
where $c_0$ and $c_1$ are just constants that depends on $m$, $a$, $E$, $l$. Solving the equation above, we have two solutions
\begin{equation}
r(\lambda) =  \pm \frac{\sqrt{c_0}\sinh(\sqrt{c_1}(\lambda \pm c_2))}{\sqrt{c_1}},
\end{equation}
where $c_2$ is a integration constant.
Once again, analyzing the above equation, we see that there are no restrictions on the geodesic, particularly for $r = 0$ \cite{Rodrigues:2023fps,Pal:2023rvv}.  From this we conclude that the function is extensible at the origin.

In the next section, we will study the types of field that can serve as sources for this geometry in the context of the $f(R)$ theory.

\section{Field Sources}\label{S:Fields}
It is well known that the SV BB is a solution from GR if we consider the coupling between a phantom scalar field and a nonlinear electrodynamics \cite{Bronnikov:2021uta,Canate:2022gpy,Rodrigues:2023vtm,Pereira:2024rtv}. However, it is not yet clear whether this type of behavior persists in modified theories of gravity. In a recent paper, the present authors have shown that this is possible for $K-$gravity theories \cite{Alencar:2024nxi}. However, the $f(R)$ case is fundamentally different since it must recover GR ($f(R)\to R$) for some parameters. 

\subsection{General Aspects}\label{general}
We assume that the SV metric represents a solution to $f(R)$ gravity theory and aim to identify the types of field that could act as sources for this geometry. We will consider, to better identify the modifications, that
\begin{eqnarray}\label{fdef}
    f(R) = a_0R + H(R).
\end{eqnarray}

We recover GR for $a_0=1$ and $H(R)=0$. For $H(R)\neq 0$, the function $f(R)$ must satisfy some viability conditions. 
The most fundamental ones are $f_R(R)=df(R)/dR > 0$, which guarantees a positive effective gravitational coupling and avoids ghost-like gravitons, and 
$f_{RR}(R)=d^2f(R)/dR^2 > 0$, which prevents the Dolgov--Kawasaki instability. Furthermore, additional information about the stability of the theory is encoded in the mass of the scalaron, given by~\cite{Shtanov:2022xew}
\begin{equation}\label{mpsi}
    m_{\psi}^{2}
    = \frac{1}{3} \left[ \frac{1}{f_{RR}(R)} - \frac{R\,}{f_R(R)} \right].
\end{equation}
The sign of $m_{\psi}^{2}$ determines whether the scalaron is located at a local minimum or maximum of the effective potential and therefore whether the background solution is stable under perturbations. Stability is achieved when $m_{\psi}^{2} > 0$, which ensures the absence of tachyonic modes.

Field sources for the linear term have also been found in many cases. Therefore, it is very appealing to consider that the matter fields will also split as $L^{(GR)}+L^{(H)}$. Therefore, we consider an action of the form:
\begin{equation}\label{action}
    S = \int \sqrt{|g|}d^4x\left[a_0 R+H(R) - 2h(\phi)g^{\mu\nu}\partial_\mu\phi\partial_\nu\phi + 2V + L(F)\right],
\end{equation}
where $H(R)$ is a function of the Ricci scalar $R = g^{\mu\nu}R_{\mu\nu}$, $\phi$ is the scalar field, $V(\phi)=V^{(GR)}(\phi)+V^{(H)}(\phi)$ is potential associated to the scalar field, $L(F)=L^{(GR)}(F)+ L^{(H)}(F)$ is the nonlinear Lagrangian, and $h(\phi)=\epsilon+h^{(H)}(\phi)$, with $h^{(H)}(\phi)$ being a correction to the phantom action. The function $h(\phi)$ modulates the kinetic contribution. When $h(\phi)>0$, the kinetic term has the conventional sign and the field is in a canonical regime, acting as a standard minimally coupled scalar. If $h(\phi)<0$, the kinetic piece acquires the opposite sign and the field enters a phantom-like regime, which tends to favor the negative kinetic energy.
However, fulfillment or violation of the various energy conditions depends on the full stress-energy tensor, including potential $V(\phi)$ and possible couplings in $L(F)$. Thus, while a negative kinetic term often facilitates violations of energy conditions, useful, for example, in wormhole or BB solutions, it is the total balance of kinetic, potential, and additional matter contributions that ultimately determines whether such conditions are satisfied or not.

Varying the action \eqref{action} with respect to $A_\mu$, $\phi$, $g^{\mu\nu}$ and splitting the stress-energy tensor, we get the following field equations
\begin{equation}\label{Eq_Max}
    \nabla_\mu(L_F F^{\mu\nu}) = \frac{1}{\sqrt{|g|}}\partial_\mu(\sqrt{|g|}L_F F^{\mu\nu}) = 0,
\end{equation}
\begin{equation}\label{Eq_scalar}
    2\epsilon\nabla_\mu\nabla^\mu\phi+ 2h^{(H)}(\phi)\nabla_\mu\nabla^\mu\phi + \frac{dh^{(H)}(\phi)}{d\phi}\partial^\mu\phi\partial_\mu\phi = - \frac{dV(\phi)}{d\phi},
\end{equation}
\begin{equation}
    a_0G_{\mu\nu}+H_{\mu\nu} = T^{(GR)}[\phi]_{\mu\nu}+T^{(H)}[\phi]_{\mu\nu} + T^{(GR)}[F]_{\mu\nu}+ T^{(H)}[F]_{\mu\nu}.\label{Eq_fR}
\end{equation}
In the last equation, we have defined the variation of $H$ as 
\begin{equation}
     H_{\mu\nu}=H_RR_{\mu\nu} - \frac{1}{2}g_{\mu\nu}H + (g_{\mu\nu}\Box - \nabla_\mu\nabla_\nu)H_R,
\end{equation}
where $L_F = dL/dF$, $H_R=dH(R)/dR$ and
\begin{align}
    T^{(GR)}[F]_{\mu\nu} &= \frac{1}{2}g_{\mu\nu}L^{(GR)}(F) - 2L^{(GR)}_FF^\alpha_\nu F_{\mu\alpha},\\
T^{(H)}[F]_{\mu\nu} &= \frac{1}{2}g_{\mu\nu}L^{(H)}(F) - 2L^{(H)}_FF^\alpha_\nu F_{\mu\alpha},\\
    T^{(GR)}[\phi]_{\mu\nu} &= 2\epsilon\partial_\mu\phi\partial_\nu\phi - g_{\mu\nu}(\epsilon\partial^\alpha\phi\partial_\alpha\phi - V^{(GR)}(\phi)),\\
     T^{(H)}[\phi]_{\mu\nu} &= 2h^{(H)}(\phi)\partial_\mu\phi\partial_\nu\phi - g_{\mu\nu}(h^{(H)}(\phi)\partial^\alpha\phi\partial_\alpha\phi - V^{(H)}(\phi)).
\end{align}
Following our approach, we will consider the following system of equations
\begin{equation}
2\epsilon\nabla_\mu\nabla^\mu\phi =- \frac{dV^{(GR)}(\phi)}{d\phi},\label{phiGR}
\end{equation}
\begin{equation}
   a_0G_{\mu\nu}= T^{(GR)}[\phi]_{\mu\nu}+T^{(GR)}[F]_{\mu\nu},\label{Eq_GR}
\end{equation}
and look for $h^{(H)}$, $L^{(H)}(F)$, and  $V^{(H)}$ such that 
\begin{equation}\label{Eq_MaxH}
    \nabla_\mu(L_F F^{\mu\nu}) = \frac{1}{\sqrt{|g|}}\partial_\mu(\sqrt{|g|}L_F F^{\mu\nu}) = 0,
\end{equation}
\begin{equation}\label{Eq_scalarH}
   2h^{(H)}(\phi)\nabla_\mu\nabla^\mu\phi + \frac{dh^{(H)}(\phi)}{d\phi}\partial^\mu\phi\partial_\mu\phi = - \frac{dV^{(H)}(\phi)}{d\phi},
\end{equation}
\begin{equation}
H_{\mu\nu} =T^{(H)}[\phi]_{\mu\nu}+ T^{(H)}[F]_{\mu\nu}\label{Eq_H},
\end{equation}
are satisfied. 

We must be careful with the possibility that the solutions for $F$ allow for the splitting of $T[F]_{\mu\nu}$. To see this, we will consider, from now on, that our space-time is spherically symmetric. By solving Maxwell's equations, equation \eqref{Eq_Max}, considering electric and magnetic charges, the non-zero components are
\beq \label{FEFB}
 F^{10}=\frac{q_e}{\Sigma(r)^2L_F}, \quad  F_{23}=q_m\sin\theta \to  F=-\frac{q_{e}^2}{2L_{F}^2\Sigma(r)^4}, \quad F=\frac{q_{m}^2}{2\Sigma(r)^4},
\eeq
where $q_e$ and $q_m$ represent, respectively, constant electric and magnetic charges. Due to $L_{F}$ in the denominator, it is impossible to split the stress-energy tensor. Therefore, our approach is consistent only for a magnetic charge, which we will call $q$ from now on. It is important to emphasize that electrically charged solutions can also be obtained. However, when considering magnetically charged sources, it becomes much simpler to separate the functions into the part corresponding to GR and the contributions from $f(R)$ gravity. For the electrically charged case, we can likewise obtain the functions corresponding to the source fields, but in general it is more complicated to make the separation between these contributions explicit. In \cite{Rodrigues:2015ayd,Rodrigues:2016fym}, the authors consider electrically charged BH solutions, and it can be seen that $L_F$ cannot be explicitly separated into a part corresponding to GR and another part corresponding to $f(R)$ gravity.

It is also interesting to note that this result does not depend on the formalism adopted. 
When attempting to obtain electric solutions using the \(P/\mathscr{H}\) formalism, we define:
\begin{equation}
    P^{\mu\nu} = L_F(F)F^{\mu\nu}, \;\; P = P_{\mu\nu}P^{\mu\nu} = L_F(F)^2F,
\end{equation}
so that, as a result, Maxwell's equations can be written as $\nabla_\mu P^{\mu\nu} = 0$. 
For a spherically symmetric configuration, this implies:
\begin{equation}
    \nabla_\mu P^{\mu\nu} = 0 \to P^{10} = \frac{q_e}{\Sigma(r)^2}.
\end{equation}
Through a Legendre transformation, we introduce the electromagnetic Hamiltonian
\begin{equation}
    \mathscr{H} = 2FL_F - L(F).
\end{equation}
However, it is straightforward to see that in the electric case it is not possible to separate the Hamiltonian as $ \mathscr{H} =  \mathscr{H}^{(GR)} +  \mathscr{H}^{(H)}$. In fact, it can also be shown that the energy–momentum tensor for the electric configuration in this formalism does not split in this way either.
As a consequence, the duality transformation does not, in general, allow one to obtain the 
electric source as a simple sum of the contribution from GR and the term 
arising from $H(R)$.

Let us review some key facts about \eqref{Eq_GR} and how the stress-energy tensor is determined by geometry. First, we note that, by combining the ${0\choose 0}$ and ${1\choose 1}$, we get 
\beq        \label{T01}
		T^{(GR)}[\phi]^{1}_{\ 1} - T^{(GR)}[\phi]^{0}_{\ 0}=- 2  \epsilon  A \phi '^2 \ =a_0 G^{1}_{\ 1} - a_0 G^{0}_{\ 0} = \frac{2 a_0 A \Sigma ''}{\Sigma},
\eeq  
and 
\begin{equation}\label{phiGR2}
 \phi '^2 = -\frac{ a_0  \Sigma ''}{\epsilon \Sigma}. 
\end{equation}
From now on, without losing generality, we will consider $a_0 = 1$. By replacing this in Eq. \eqref{phiGR}, we find that $\phi(r)$ and $V^{(GR)}$ and therefore  $T^{(GR)}[\phi]_{\mu\nu}$ are determined from the components ${0\choose 0}$ and ${1\choose 1}$ of the Einstein tensor. With this, we write \eqref{Eq_GR} as
\begin{equation}
T^{(GR)}[F]_{\mu\nu}= G_{\mu\nu}- T^{(GR)}[\phi]_{\mu\nu}.
\end{equation}
From the ${0\choose 0}$ and by combining ${0\choose 0}$ and ${2\choose 2}$ components of the above equation, we get
\begin{equation}
\label{LGR}L^{(GR)}[F]= G^{0}_{\ 0}- T^{(GR)}[\phi]^{0}_{\ 0},
\end{equation}
\begin{equation}
\frac{q_{}^2}{2\Sigma(r)^4}L_F^{(GR)}[F]= \left(G^{0}_{\ 0}-G^{2}_{\ 2}\right)- \left(T^{(GR)}[\phi]^{0}_{\ 0}-T^{(GR)}[\phi]^{2}_{\ 2}\right).
\end{equation}
Therefore, as usual, $\phi(r)$, $V^{(GR)}$, $L(r)^{(GR)}$, and $L_F(r)^{(GR)}$ are determined from the geometry. The explicit dependence on the fields will depend on a case-by-case study. 

Now we will show that, given the above solution, we can also determine $T^{(H)}[\phi]_{\mu\nu}$ and $T^{(H)}[F]_{\mu\nu}$ from geometry. For this, we come back to Eq. (\ref{Eq_H}) and note that $T^{(H)}[\phi]_{\mu\nu}$ and $T^{(H)}[F]_{\mu\nu}$ has exactly the same symmetries as $T^{(GR)}[\phi]_{\mu\nu}$ and $T^{(GR)}[F]_{\mu\nu}$ respectively. Therefore, the same steps can be followed, and we get directly 
\beq        \label{TphiH}
		T^{(H)}[\phi]^{1}_{\ 1} - T^{(H)}[\phi]^{0}_{\ 0}=- 2  h^{(H)}  A \phi '^2 \ =2  A h^{(H)} \frac{ \Sigma ''}{\epsilon \Sigma}=H^{1}_{\ 1} - H^{0}_{\ 0}.
\eeq  
By replacing this at Eq. (\ref{Eq_scalarH}), we find that $h^{(H)}, V^{(H)}$, and $T^{(H)}[\phi]_{\mu\nu}$ are completely determined. Just as before, these results can be used to obtain
\begin{equation}\label{LH}
L^{(H)}[F]= H^{0}_{\ 0}- T^{(H)}[\phi]^{0}_{\ 0},
\end{equation}
\begin{equation}\label{LFH}
\frac{q_{}^2}{2\Sigma(r)^4}L_F^{(H)}[F]= \left(H^{0}_{\ 0}-H^{2}_{\ 2}\right)- \left(T^{(H)}[\phi]^{0}_{\ 0}-T^{(H)}[\phi]^{2}_{\ 2}\right).
\end{equation}

The above results can be further simplified if we note that 
\begin{equation}
    H_{\mu\nu} = H_R\left(T^{(GR)}_{\mu\nu}-\frac{1}{2}g_{\mu\nu}(T^{(GR)})_{\alpha}^{\alpha}\right) - \frac{1}{2}g_{\mu\nu}H + (g_{\mu\nu}\Box - \nabla_\mu\nabla_\nu)H_R.
\end{equation}
Replacing this at (\ref{TphiH}), (\ref{LH}), and (\ref{LFH}), we get the final source
\begin{equation}
    2  A h^{(H)} \frac{ \Sigma ''}{\epsilon \Sigma}=H_R\left((T^{(GR)})^{1}_{1}-(T^{(GR)})^{0}_{0}\right) -(\nabla^1\nabla_1-\nabla^0\nabla_0)H_R,
\end{equation}
\begin{equation}
L^{(H)}[F]= H_R\left((T^{(GR)})^{0}_{\ 0}-\frac{1}{2}(T^{(GR)})_{\alpha}^{\alpha}\right) - \frac{1}{2}H + (\Box - \nabla^0\nabla_0)H_R- T^{(H)}[\phi]^{0}_{\ 0},
\end{equation}
\begin{equation}
\frac{q_{}^2}{2\Sigma(r)^4}L_F^{(H)}[F]= H_R\left((T^{(GR)})^{0}_{0}-(T^{(GR)})^{2}_{2}\right) -(\nabla^0\nabla_0-\nabla^2\nabla_2)H_R- \left(T^{(H)}[\phi]^{0}_{\ 0}-T^{(H)}[\phi]^{2}_{\ 2}\right).
\end{equation}
Therefore, our approach can be used to find the source to any $f(R)$ theory defined by (\ref{fdef}).

Given the form of the equations \eqref{Eq_fR} and \eqref{Eq_GR}, considering the way we constructed our action, it is easy to see that our functions of interest will take the form
\begin{eqnarray}
     F(r)&=&\frac{2 q^2}{\left(a^2+r^2\right)^2},\\
       L(F) &=& L^{(GR)}(F) + L^{(H)}(F),\\
     L_F(F) &=&   L_F^{(GR)}(F) + L_F^{(H)}(F),\\
     \phi(r) &=& \arctan\left(\frac{r}{a}\right),\\
\label{phitotal}     h(\phi) &=& h^{(GR)}(\phi) + h^{(H)}(\phi),\\
     V(\phi) &=& V^{(GR)}(\phi) + V^{(H)}(\phi),
\end{eqnarray}
where the functions of GR are given by\cite{Bronnikov:2021uta}
\begin{eqnarray}
    L^{(GR)}(F(r)) &=& \frac{12 a^2 m}{5 \left(a^2+r^2\right)^{5/2}},\\
    L_F^{(GR)}(F(r)) &=& \frac{3 a^2 m}{2 q^2 \sqrt{a^2+r^2}},\\
    h^{(GR)}(\phi(r))&=& \epsilon = -1,\\
    V^{(GR)}(\phi(r)) &=& \frac{4 a^2 m}{5 \left(a^2+r^2\right)^{5/2}}.
\end{eqnarray}
Or, in terms of the electromagnetic invariant $F$ and the scalar field $\phi$.
\begin{eqnarray}
      L^{(GR)}(F) &=& \frac{3\ 2^{3/4} a^2 F^{5/4} m}{5 q^{5/2}},\\
    h^{(GR)}(\phi)&=& \epsilon = -1,\\
    V^{(GR)}(\phi) &=& \frac{4 m \cos ^5\phi }{5 a^3}.
\end{eqnarray}

We will now divide our analysis into different cases, i.e., various models of $f(R)$ theories.

\subsection{Case I: \texorpdfstring{$H(R) = a_RR^2$}{}}

The most well-known $f(R)$ theory model is the so-called Starobinsky model \cite{Starobinsky:1980te,DeFelice:2010aj,Nojiri:2010wj,Nojiri:2017ncd}, where
\begin{equation}
    f(R) = R + a_R R^2,
\end{equation}
where $a_R$ is a coupling constant. In this context, $H(R) = a_RR^2$, and GR is recover for $H = 0$, i.e., $a_R = 0$.
The function $H_R$ is consequently given by $H_R = 2a_R R$.
Using the metric given in \eqref{line} and following the approach presented in \ref{general}, we can analytically determine all the correction functions of interest, that are given by:
\begin{eqnarray}
    L^{(H)}(F(r))&=& a_R a^2\left(\frac{2 \left(20 a^4+a^2 \left(27 m^2+100 r^2\right)+675 m^2 r^2+80 r^4\right)}{15 \left(a^2+r^2\right)^5}-\frac{16  m \left(11 a^2+81 r^2\right)}{21 \left(a^2+r^2\right)^{9/2}}\right),\\
    L_F^{(H)}(F(r))&=&\frac{6 a^2 a_R m \left(a^2-9 r^2\right)}{q^2 \left(a^2+r^2\right)^{5/2}}+\frac{2 a^2 a_R \left(a^2 \left(4 r^2-9 m^2\right)+45 m^2 r^2+4 r^4\right)}{q^2 \left(a^2+r^2\right)^3},\\
    h^{(H)}(\phi (r))&=&\frac{2 a_R \left(a^2-10 r^2\right) \left(2 \sqrt{a^2+r^2}-9 m\right)}{\left(a^2+r^2\right)^{5/2}},\label{hphi_model1}\\
    V^{(H)}(\phi (r))&=& a_Ra^2  \left(\frac{2 m \left(1290 r^2-481 a^2\right)}{21
   \left(a^2+r^2\right)^{9/2}}+\frac{130 a^4+a^2 \left(783 m^2-70 r^2\right)-25 r^2 \left(135 m^2+8 r^2\right)}{15
   \left(a^2+r^2\right)^5}\right).
\end{eqnarray}
These terms involving $a_R$ are the corrections required by the modified gravity theory for the sources so that the SV metric can be generated by the theory.

It is important to highlight some differences from GR. In GR, we have $h(\phi) = -1$, so that the scalar field is always a phantom field. Here, using equation \eqref{phitotal}, considering the correction term find above, we can expand $h(\phi)$
\begin{eqnarray}
    h(\phi)&\approx& -1-\frac{40a_R}{r^2}+O\left(r^{-3}\right), \quad \mbox{if} \quad r\rightarrow \infty,\\
    h(\phi)&\approx& -1+\frac{2 a_R (2 a-9 m)}{a^3}+O\left(r^{2}\right), \quad \mbox{if} \quad r\rightarrow 0.
\end{eqnarray}
These expressions tell us that the scalar field continues to exhibit phantom behavior at infinity. However, if $a > \frac{9m}{2}$, that's, in the traversable wormhole case, and $a_R$ is sufficiently large and positive, the scalar field may cease to be phantom in more internal regions, thus becoming partially phantom. In this way, the $f(R)$ theory can relax the condition of the scalar field being phantom in some regions of space-time.

The functions $F(r)$ and $\phi(r)$ are easily invertible so that we can analytically write $L^{(H)}(F)$, $V^{(H)}(\phi)$, and $h^{(H)}(\phi)$, which are expressed as:
\begin{eqnarray}
    L^{(H)}(F)&=&\frac{a^2 a_R F^{5/4}}{210 \sqrt{2} q^5} \left[5 q \left(945 \sqrt{2} F^{3/4} m^2-648\ 2^{3/4} m \sqrt{F q}+224 \sqrt[4]{F} q\right)\right.\\\nonumber
    &-&\left.28 a^2 \left(162 F^{5/4} m^2+15 \sqrt{2} F^{3/4} q-100  \sqrt[4]{2} F m \sqrt{q}\right)\right],\\
    h^{(H)}(\phi)&=&\frac{a_R (11 \cos 2 \phi-9) \left(2a \sqrt{ \sec ^2\phi}-9 m\right)}{\left(a^2 \sec ^2\phi\right)^{3/2}},\\
    V^{(H)}(\phi)&=&\frac{a_R \cos ^6\phi}{105 a^6} \left(m \cos ^2\phi \left(77 \cos 2 \phi \left(189 m-115 a\sec \phi\right)+4045 a\sec\phi-9072 m\right)
    +35 a^2
   (33 \cos 2 \phi-7)\right).
\end{eqnarray}
The presence of $f(R)$ theory requires much more nonlinearities in the field sources than GR.

It is important to emphasize that the electromagnetic functions found here satisfy the consistency relation given by:  
\begin{equation}  
L_F \left( \frac{dF}{dr} \right) - \frac{dL(F)}{dr} = 0.\label{eq_cons}  
\end{equation}  

Thus, we see that, even with the additional complications that $f(R)$ theory demands from the source fields, it is possible to obtain the SV solution in modified theories of gravity by combining a nonlinear electrodynamics with a partially phantom scalar field.

We can also choose other $f(R)$ models and verify what types of sources emerge. However, in the next subsection we will attempt a different approach.  

\subsubsection{A brief comment about viability conditions and scalaron mass for case I}

The mass of the scalaron in the Starobinsky model has been extensively studied in the literature. This model satisfies the classical and quantum stability conditions, requiring that $f_R(R) > 0$ and $f_{RR}(R) > 0$, which ensures a positive and well-behaved scalaron mass \cite{appleby2009curing,gannouji2012generic,gorbunov2014scalaron}. Furthermore, these works demonstrate that the model avoids curvature singularities and maintains a viable cosmological evolution \cite{gannouji2012quantum,aldabergenov2018beyond}.

\subsection{Case II: \texorpdfstring{$H_R(R(r)) =  a_1r$}{}}
In references \cite{Rodrigues:2015ayd,Rodrigues:2016fym}, the authors studied regular BHs in $f(R)$ theory and found that, due to the symmetry of the space-time they considered, the function $f_R(R)$ was linear when analyzed in terms of the radial coordinate. Since we are considering BB, which are more complex structures than usual regular BHs, this type of behavior does not naturally arise from the field equations. However, we can still impose such behavior and investigate what types of field sources may emerge.  

Let us consider the $H_R$ written as
\begin{equation}\label{fr1}
    H_R(R(r)) = a_1 r,
\end{equation}
that is, $f_R = 1 + a_1r$, where $a_1$ is a constant. In this case, the function $H(R)$ as a function of the radial coordinate can be calculated as:  
\begin{equation}\label{f_rcase2}
    H(R(r))=\int H_R\frac{dR}{dr}dr.
\end{equation}
 When $a_1=0$ we recover GR.

 Considering the line element \eqref{line} and following the approach presented in \ref{general}, we find that the functions related to the field sources corrections are given by:
 \begin{eqnarray}
    L^{(H)}(F(r))&=&2 a_1 \left(-\frac{2 m \left(10 a^2 r^3+7 r^5\right)}{5 a^2
   \left(a^2+r^2\right)^{5/2}}-\frac{r}{a^2+r^2}+\frac{2 \tan ^{-1}\left(\frac{r-\sqrt{a^2+r^2}}{a}\right)}{a}\right),\\
   L_F^{(H)}(F(r))&=&\frac{a_1 }{2 q^2}\left(-r \sqrt{a^2+r^2} \left(\frac{3 m r^2}{a^2+r^2}-6
   m\right)-r \left(a^2+r^2\right)\right),\\
   h^{(H)}(\phi (r))&=&-a_1 r,\qquad \phi(r)= \arctan\left(\frac{r}{a}\right),\\
   V^{(H)}(\phi (r))&=&a_1 \left(\frac{\tan
   ^{-1}\left(\frac{\sqrt{a^2+r^2}-r}{a}\right)}{a}+\frac{2 m r \left(5 a^4+5 a^2 r^2+2 r^4\right)}{5 a^2
   \left(a^2+r^2\right)^{5/2}}-\frac{r}{2 \left(a^2+r^2\right)}\right).
 \end{eqnarray}
 In this case, unlike the previous model, since, by equation \eqref{phitotal}, $h(\phi(r)) = -1 - a_1r$, the scalar field will always be phantom, since $h(\phi)$ will always be negative if we consider $a_1$ as a positive constant. The scalar field will be canonic in some regions only if $a_1$ assumes negative values.

It is possible to invert the functional dependencies once more, expressing the quantities $L^{(H)}$, $V^{(H)}$, and $h^{(H)}$ as functions of $F$ and $\phi$, respectively.
\begin{eqnarray}
   L^{(H)}(F) &=& 
a_1 \left( 
    - \sqrt{2\sqrt{\frac{2F}{q^2}} - \frac{2a^2F}{q^2}}\right.
    - 
    \frac{2^{3/4} m \left( \sqrt{\frac{2q^2}{F}} -a^2\right)^{3/2} 
    \left(3 a^2 + 7 \sqrt{\frac{2q^2}{F}}\right)F^{5/4}}{5 a^2q^{5/2}} \nonumber\\
    &+& \left.
    \frac{4}{a}\arctan\left(\frac{-2^{1/4} q^{1/2}F^{-1/4} + 
    \sqrt{-a^2 + \sqrt{2} |q|F^{-1/2} }}{a}\right) 
\right),\\
   h^{(H)}(\phi) &=&  - a_1a\tan\phi,\\
V^{(H)}(\phi)&=&  
\frac{a_1}{20 a^2} \left(20a \arctan\left( \sec\phi - \tan\phi \right) 
- 5a \sin(2\phi)+
2 m \left(13 + 6 \cos(2\phi) + \cos(4\phi)\right) \sin\phi\right).
\end{eqnarray}

\begin{figure}
    \centering
    \includegraphics[width=1\linewidth]{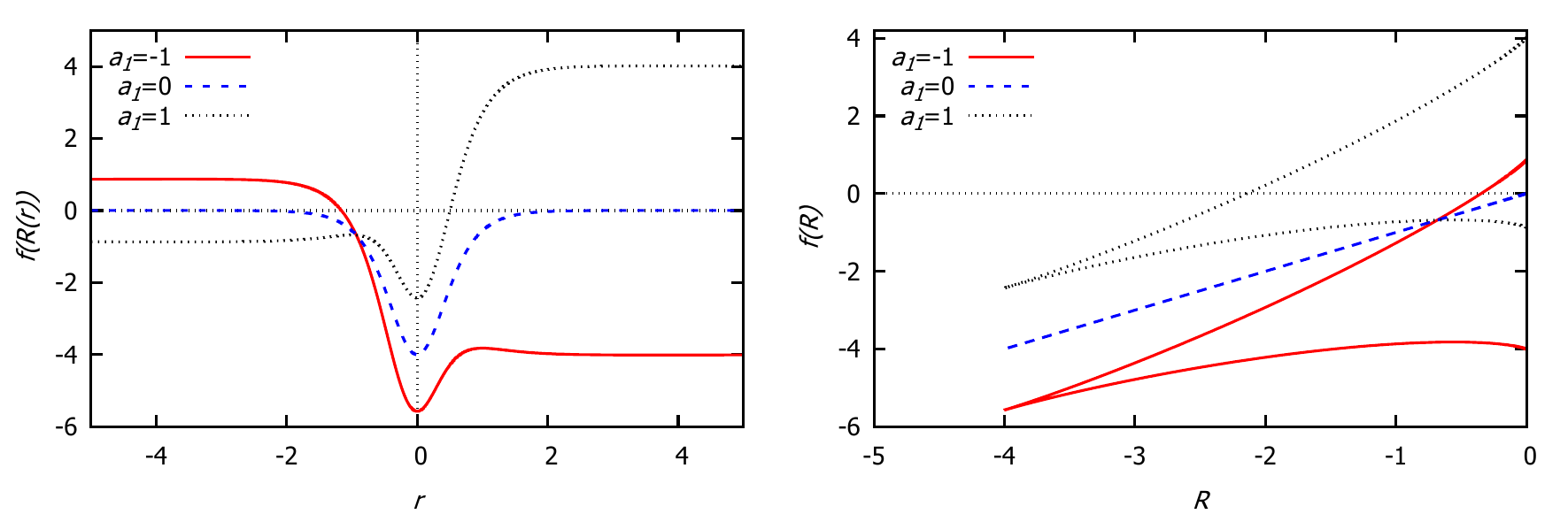}
    \caption{Behavior of $f(R)$ as a function of the radial coordinate (left panel) and as a function of the curvature scalar (right panel), fixing $a = m = 1$ and varying the values of the parameter $a_1$. As we can see, $f(R(r))$ is not symmetric under the transformation $r \to -r$ when $a_1 \neq 0$. This happens because $f_R(r)$ is linear in the radial coordinate, $f_R = 1 + a_1 r$. Nevertheless, the curvature scalar is symmetric under $r \to -r$. This means that under this transformation we obtain the same value of the curvature scalar but a different value of $f(R(r))$, leading to the multivalued behavior shown in the right-hand figure: the same value of the curvature scalar can correspond to two different values of $f(R)$.}
    \label{fig:fR_fRr_model2}
\end{figure}

Once again, we observe that the presence of the $H(R)$ function induces greater nonlinearity in the obtained fields.  
Integrating the equation \eqref{f_rcase2}, we obtain
\begin{equation}
    H(R(r)) =  
 \frac{a_1mr^3\left(10  a^2+4  r^2\right)}{a^2(a^2 + r^2)^{5/2}} 
+ \frac{a^2 a_1 r}{(a^2 + r^2)^2} 
- \frac{a_1 r^3}{(a^2 + r^2)^2} 
- \frac{2 a_1 \arctan\left(\frac{r - \sqrt{a^2 + r^2}}{a}\right)}{a}.
\end{equation}
However, in this case, analytically inverting $R(r)$ in equation \eqref{RicciScalar} for the Ricci scalar is not possible, so we cannot write an analytical form for the function $H(R)$. To generate the plot, we recall that the function $f(R)$ is given by $f(R) = R + H(R)$. Nevertheless, in Fig. \ref{fig:fR_fRr_model2}, we observe the behavior of the function $f(R)$. It is clear that the only symmetric case under the exchange $r \to -r$ occurs for $a_1 = 0$, which corresponds exactly to the GR case. For $a_1 = 0$, the function $f(R)$ is just a straight line, as expected, while for other cases, we have a multivalued function with different curves. This multivalued behavior arises precisely due to the asymmetry in the transformation $r \to -r$ in the function $H(R(r))$.

\subsubsection{A brief comment about viability conditions and scalaron mass for case II}

The first viability condition requires that $f_R > 0$, which in this case is trivially satisfied in the regions where $r > -1/a_1$. To analyze the second condition, where $f_{RR} > 0$, we first note that

\begin{equation}
    f_{RR}(R(r)) = -\frac{a_1 \left(a^2+r^2\right)^{5/2}}{2 a^2 \left(\sqrt{a^2+r^2}-3 m\right)}.
\end{equation}
The sign analysis of $f_{RR}(R(r))$ can be performed by noting that the factors $(a^2+r^2)^{5/2}$ and $2a^2$ are always positive, so that
\begin{equation}
\operatorname{sign}\!\big(f_{RR}(R(r))\big)
=
-\,\operatorname{sign}(a_1)\,
\operatorname{sign}\!\big(\sqrt{a^2+r^2}-3m\big).
\end{equation}
It is convenient to define $r_c=\sqrt{9m^2-a^2}$ for $a<3m$, since the denominator of $f_{RR}(R(r))$ vanishes at $r=\pm r_c$, producing singularities in function $f_{RR}$. In this case, the term $\sqrt{a^2+r^2}-3m$ is negative for $|r|<r_c$ and positive for $|r|>r_c$, so the sign of $f_{RR}$ flips across these points. When $\operatorname{sign}(a_1)=+1$, the physical viability condition $f_{RR}>0$ is satisfied only inside the interval $-r_c<r<r_c$ (excluding $r=\pm r_c$), whereas for $\operatorname{sign}(a_1)=-1$ the allowed region lies outside, that is, $|r|>r_c$. If instead $a>3m$, the quantity $\sqrt{a^2+r^2}-3m$ never changes sign: in the limiting scenario $3m=a$, the point $r=0$ becomes singular and is removed from the analysis, leaving no admissible region for $a_1>0$ and all $r\neq0$ allowed when $a_1<0$. Finally, if $a>3m$, the denominator remains positive everywhere, implying that $f_{RR}>0$ only when $a_1<0$, while for $a_1>0$ the function $f_{RR}$ is negative throughout the real line.
In summary, the conditions for physical viability translate into the following concise classification: for $3m>a$ the theory is allowed either in the interior region $|r|<r_c$ if $a_1>0$, or only in the asymptotic exterior region $|r|>r_c$ if $a_1<0$; when $3m=a$, viability holds for all $r\neq0$ provided $a_1<0$ and there are no permitted regions for $a_1>0$; and when $3m<a$, the full real axis is admissible if $a_1<0$, while no physically acceptable configuration exists for $a_1>0$.

\begin{figure}[ht]
    \centering
    \includegraphics[width=0.5\linewidth]{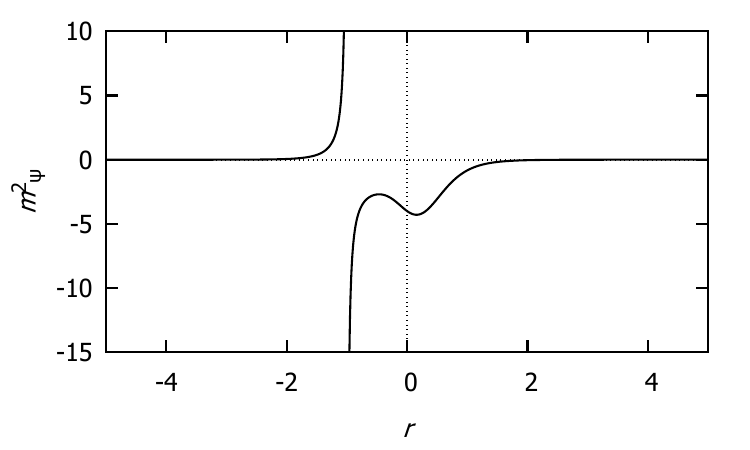}\includegraphics[width=0.5\linewidth]{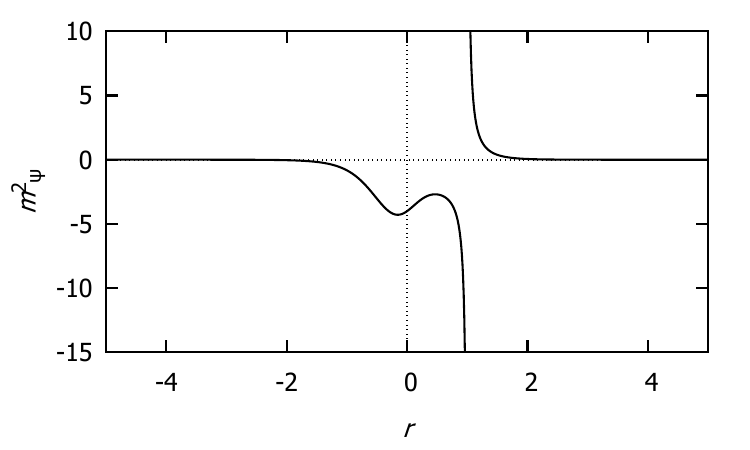}
    \caption{Scalaron mass $m_\psi^2$ as a function of the radial coordinate $r$ for $a_1 = 1$ (left) and $a_1 = -1$ (right), with $a = m = 1$. Again, for both \( a_1 > 0 \) and \( a_1 < 0 \), there are regions where the scalaron mass is positive.} 
    \label{fig:mpsi2}
\end{figure}

Using  the equation \eqref{mpsi}, we can write the scalaron mass for this case as
\begin{equation}
m_{\psi}^{2}(R(r)) = \frac{2 a^2 \left(3 a^2 a_1 \left(\sqrt{a^2+r^2}-3 m\right)+r \sqrt{a^2+r^2} (7 a_1 r+4)-3 m r (8 a_1 r+5)\right)}{3 a_1 \left(a^2+r^2\right)^{7/2} (a_1
   r+1)}.
\end{equation}
Now, we analyze the behavior of $m^2_\psi$ graphically. In Fig.~\ref{fig:mpsi2}, we show the plot of the scalaron mass as a function of the radial coordinate for $a_1 = 1$ (left) and $a_1 = -1$ (right). From the graph, we see once again that for both parameter values there exist regions where $m^2_{\psi} > 0$, indicating that it is always possible for locally stable regions to exist in this case. It is also interesting to note that, since $H_R$ is not symmetric under the transformation $r \to -r$, the graph is also not symmetric under this transformation.

\subsection{Case III: \texorpdfstring{$H_R (R(r)) =  a_2r^2$}{}}
For case II, we see that the behavior of $f(R(r))$ changes if we make the transformation $r \to -r$. This fact occur because the function $H_R$ is not symmetric under transformation $r \to -r$. Moreover, in the reference \cite{Rodrigues:2015ayd}, given a metric of the form
\begin{equation}
    ds^2 = A(r)dt^2 - B^{-1}(r)dr^2 - r^2d\Omega^2,
\end{equation}
the condition \eqref{fr1} is obtained by imposing $A(r) = B(r)$. However, in BB metrics, such as \eqref{line}, we can always make a coordinate transformation given by
\begin{equation}
  \tilde{r}^2 = \Sigma(r)^2,   
\end{equation}
such that the metric can be written in the form
\begin{equation}
    ds^2 = U(\tilde{r})dt^2 - G^{-1}(\tilde{r})d\tilde{r}^2 - \tilde{r}^2d\Omega^2,
\end{equation}
where $U(\tilde{r}) \neq G(\tilde{r})$. In this case, since the function $H_R(R(r))$ does not necessarily need to retain the specific form \eqref{f_rcase2}, we can propose alternative formulations for it such that it remains invariant under transformation $r \to -r$. Accordingly, in this subsection, we focus on the case where
\begin{equation}
    H_R(R(r)) =  a_2r^2.
\end{equation}

Considering this function, the SV metric, and again following the approach presented in \ref{general}, we find that the functions related to the field sources corrections are now given by
\begin{eqnarray}
L^{(H)}(F(r)) &=&  \frac{(88a^4 a_2 m)}{5\left(a^2 + r^2\right)^{5/2}} + \frac{44a^2 a_2 m r^2}{\left(a^2 + r^2\right)^{5/2}} + \frac{24a_2 m r^4}{\left(a^2 + r^2\right)^{5/2}} - \frac{a_2(6a^4 +16a^2  r^2+10 r^4)}{\left(a^2 + r^2\right)^2}
+ 4a_2 \ln\left(a^2 + r^2\right)
,\\
L_F^{(H)}(F(r)) &=& \frac{-2 a_2 r^4 \left( -3 m + \sqrt{a^2 + r^2} \right) + a^2   a_2 r^2 \left( 9 m - 2 \sqrt{a^2 + r^2} \right) }{2 q^2 \sqrt{a^2 + r^2}},\\
h^{(H)}(\phi (r)) &=&  - \frac{a_2 \left( a^4 + 3 a^2 r^2 + r^4 \right)}{a^2},\\
V^{(H)}(\phi (r)) &=& \frac{a^2 a_2}{a^2 + r^2} - \frac{4m \left(6a^4 a_2 + 5a_2 r^4  + 10a^2a_2r^2 \right)}{5\left(a^2 + r^2\right)^{5/2}} - 2a_2 \ln\left(a^2 + r^2\right).
\end{eqnarray}

We can once again explicitly write $L^{(H)}$ as a function of $F$, and $V^{(H)}$ and $h^{(H)}$ as functions of $\phi$, which are
\begin{eqnarray}
L^{(H)}(F) &=& \frac{a_2}{5 q^2 \sqrt{\frac{q}{\sqrt{F}}}}\left(-3 \, 2^{3/4} a^4 F m + 10 q^2 \left(6 \, 2^{3/4} m - 5 \sqrt{\frac{q}{\sqrt{F}}}\right) \right.\nonumber\\
&+& 
\left.10 \, 2^{1/4} a^2 \sqrt{F} q \left(-m + 2^{1/4} \sqrt{\frac{q}{\sqrt{F}}}\right) + 
20 q^2 \sqrt{\frac{q}{\sqrt{F}}} \log\left(\frac{\sqrt{2} q}{\sqrt{F}}\right)\right),\\
 h^{(H)}(\phi) &=&  a^2 a_2 \left[1 - \sec^2\phi \left(1 + \sec^2\phi\right)\right],\\
  V^{(H)}(\phi) &=&  a_2 \left( \cos^2\phi - 2 \log\left(a^2 \sec^2\phi\right) - \frac{(20 m + 4m\cos^4\phi)\cos\phi }{5a}\right).
\end{eqnarray}

It is interesting to note that, for $r <<1$, we have
\begin{equation}
    h(\phi) \approx -1 - a_2a^2.
\end{equation}
This suggests that the regularization constant $a$ may also interfere with the change in the type of scalar field near the origin if $a_2$ is negative. Thus, depending on the sign of the parameter $a_2$ and the values of $a$, we can have a phantom scalar field, a partially phantom field, or a canonical scalar field.

\begin{figure}
    \centering
    \includegraphics[width=1\linewidth]{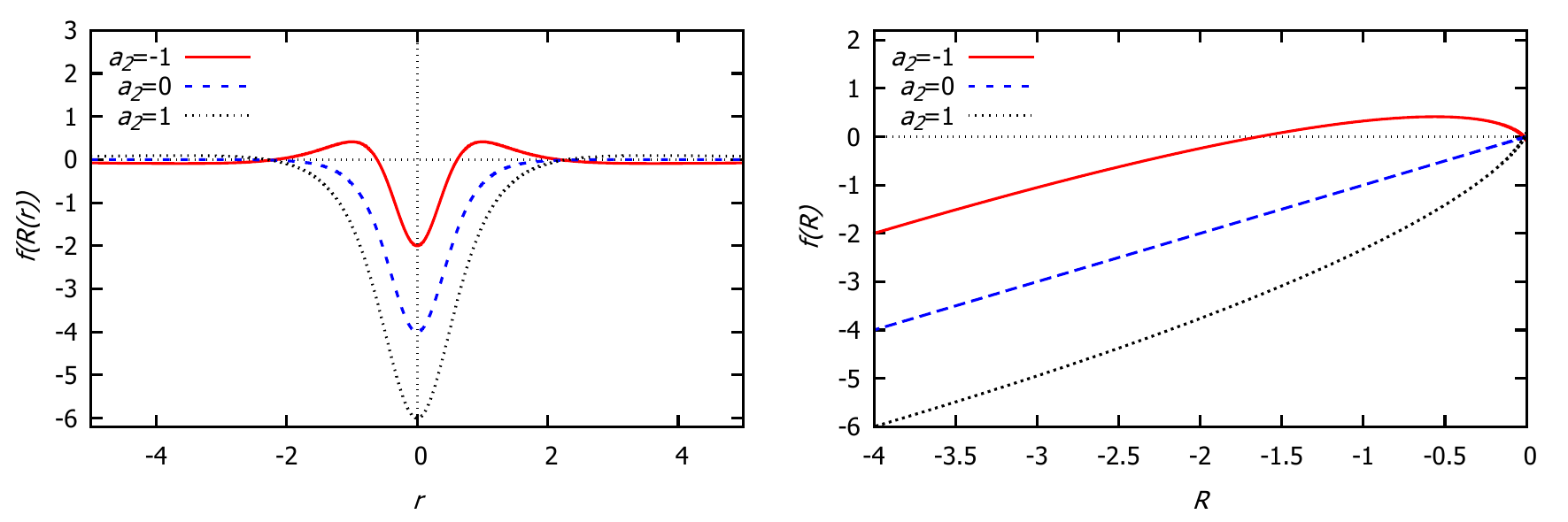}
    \caption{Behavior of $f(R)$ as a function of the radial coordinate (left panel) and as a function of the curvature scalar (right panel), fixing $a = m = 1$ and varying the values of the parameter $a_2$. In this case, the function $f(R(r))$ is symmetric under the transformation $r \to -r$, even when $a_2 \neq 0$. This ensures that we obtain a single value of $f(R)$ for each value of the curvature scalar, thereby avoiding any multivalued behavior.}
    \label{fig:fR_fRr_model3}
\end{figure}

Integrating again the equation \eqref{f_rcase2}, we obtain
\begin{equation}
    H(R(r)) = \frac{2 a^2 \left(  a^2 a_2 \left(-2 m + \sqrt{a^2 + r^2}\right) + a_2 r^2 \left(-5 m + 2 \sqrt{a^2 + r^2}\right)\right)}{\left(a^2 + r^2\right)^{5/2}}.
\end{equation}
Analyzing the equation above, we clearly see that $H(R(r))$ is symmetric under transformation $r \to -r$. By making the parametric plot of $f(r)$ vs $R(r)$, presented in Fig. \ref{fig:fR_fRr_model3}, we also observe that it preserves this symmetry, unlike the previous case.

The results obtained can be generalized by considering $H_R = a_n r^n$. However, not all values of $n$ will satisfy the electromagnetic consistency relation \eqref{eq_cons}.

\subsubsection{A brief comment about viability conditions and scalaron mass for case III}
In this case, since $f_R(R(r)) = 1 + a_2 r^2$, the condition $f_{RR} > 0$ is satisfied in all regions if $a_2 > 0$. If $a_2 < 0$, this condition is only satisfied in the regions where $|r| < 1/\sqrt{-a_2}$. For the second condition, we write

\begin{equation}
    f_{RR}(R(r)) = -\frac{a_2 r \left(a^2+r^2\right)^{5/2}}{a^2 \left(\sqrt{a^2+r^2}-3 m\right)}.
\end{equation}
The sign analysis of $f_{RR}$ can be performed by noting that the factors $(a^{2}+r^{2})^{5/2}$ and $a^{2}$ are always positive, such that 
\begin{equation}
\mathrm{sign}\!\left(f_{RR}(R(r))\right)=-\,\mathrm{sign}(a_2)\,\mathrm{sign}(r)\,\mathrm{sign}\!\left(\sqrt{a^{2}+r^{2}}-3m\right).    
\end{equation}
 Introducing $r_{c}=\sqrt{9m^{2}-a^{2}}$ whenever $a<3m$, we have $\mathrm{sign}\!\left(\sqrt{a^{2}+r^{2}}-3m\right)=-1$ for $|r|<r_{c}$ and $+1$ for $|r|>r_{c}$, while $r=\pm r_{c}$ must be excluded because they produce a divergence in the denominator. Therefore, if $\mathrm{sign}(a_2)=+1$, the physical viability condition $f_{RR}(R(r))>0$ requires opposite signs for $r$ and $\sqrt{a^{2}+r^{2}}-3m$, which leads to physically allowed regions $0<r<r_{c}$ and $r<-r_{c}$ (or equivalently $r<0$ when $a \ge 3m$). On the other hand, if $\mathrm{sign}(a_2)=-1$, the condition is reversed and both signs must coincide, producing the allowed regions $-r_{c}<r<0$ and $r>r_{c}$ (or $r>0$ when $ a \ge 3m3$). In all cases, $f_{RR}(R(r))=0$ only at $r=0$, and the points $r=\pm r_{c}$ correspond to singularities that must be excluded from the analysis. In summary, the physical viability condition $f_{RR}(R(r))>0$ imposes opposite signs between $r$ and $\sqrt{a^2+r^2}-3m$ when $a_2>0$, which restricts the allowed regions to $0<r<r_c$ and $r<-r_c$ (equivalently $r>0$ when $a\ge 3m$). Conversely, if $a_2<0$, both quantities must share the same sign, producing the viable domains $-r_c<r<0$ and $r>r_c$ (or $r>0$ for $3m\le a$). In all cases, $f_{RR}(R(r))=0$ only at $r=0$, and the points $r=\pm r_c$ correspond to singularities that must be removed from the analysis.

\begin{figure}[ht]
    \centering
    \includegraphics[width=0.5\linewidth]{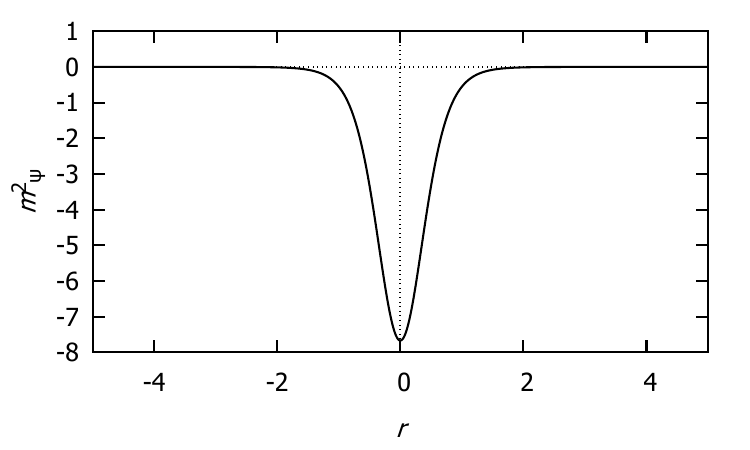}\includegraphics[width=0.5\linewidth]{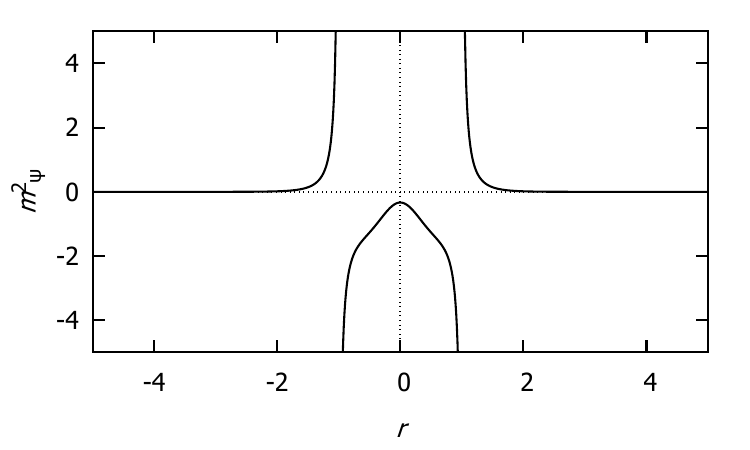}
    \caption{Scalaron mass $m_\psi^2$ as a function of the radial coordinate $r$ for $a_2 = 1$ (left) and $a_2 = -1$ (right), with $a = m = 1$. In this case, $m_\psi^2 <0$ in more central regions
 for $a_2>0$, and positive in some regions for $a_2<0$, indicating that stability regions exist for this case.}
    \label{fig:m2psi3}
\end{figure}

The application of equation~\eqref{mpsi} to this case leads to
\begin{equation}
  m_{\psi}^{2}(R(r)) =  \frac{a^2 \left(2 \sqrt{a^2+r^2} \left(3 a^2 a_2+5 a_2 r^2+2\right)-3 m \left(6 a^2 a_2+11 a_2 r^2+5\right)\right)}{3 a_2 \left(a^2+r^2\right)^{7/2}
   \left(a_2 r^2+1\right)}.\label{scalaronr2}
\end{equation}
In Fig.~\ref{fig:m2psi3} we show the plot of the scalaron mass for $H_R =a_2 r^2$ considering $a_2 = 1$ (left) and $a_2 = -1$ (right). For $a_2 > 0$, we have $m^2_{\psi} \leq 0$ in more central regions, indicating that no local stability exists in this region. However, from Eq. \eqref{scalaronr2} one sees that the negative term grows as $r^2$, whereas the positive term grows as $r^3$ for large $r$; hence, for sufficiently large $r$, 
the expression becomes positive. For $a_2 < 0$, however, the scalaron mass takes positive values in regions where $|r| >1$, for these parameters values.

\subsection{Case IV: \texorpdfstring{$H_R(R(r)) = a_{\Sigma}\Sigma$}{}}
We can test another form of $H_R$. Considering that BB typically arises through the substitution $r^2 \to r^2 + a^2$ and that for regular BHs we had $H_R(R(r)) = a_1 r$, we can apply this substitution to obtain $H_R(R(r)) = a_\Sigma \Sigma$. This model will also be symmetric under the transformation $r \to -r$.

Considering again that our space-time is described by the SV metric and following the approach presented in \ref{general}, we find that the correction functions related to the matter sources are given by:  
\begin{eqnarray}
L^{(H)}(F(r)) &=&\frac{2 a_\Sigma \left(9 m \sqrt{a^2+r^2}-4 a^2-6
   r^2\right)}{3 \left(a^2+r^2\right)^{3/2}},\\
L_F^{(H)}(F(r)) &=& \frac{a_\Sigma \left(3 a^2 m-r^2 \left(\sqrt{a^2+r^2}-3
   m\right)\right)}{2 q^2},\\
h^{(H)}(\phi (r)) &=&  -\frac{3}{2} a_\Sigma \sqrt{a^2+r^2}
,\\
V^{(H)}(\phi (r)) &=& \frac{3 a^2 a_\Sigma m}{4
   \left(a^2+r^2\right)^2}+  \frac{ a^2 a_\Sigma}{2\left(a^2+r^2\right)^{3/2}}.
\end{eqnarray}
As in the other cases, it is only possible to have a scalar field that is canonical in some regions, or partially phantom, if the constant related to the nonlinear terms of the gravitational theory, $a_\Sigma$, assumes negative values. Depending on the values of $a_\Sigma$ and $a$, it is possible to have $h(\phi)$ always positive.

We can also write $L^{(H)}(F)$, $V^{(H)}(\phi)$, and $h^{(H)}(\phi)$, which are given by:
\begin{eqnarray}
L^{(H)}(F) &=& \frac{a_\Sigma \left(10
   \sqrt{2} a^2 \sqrt{F} q+15 q^2 \left(\frac{3\ 2^{3/4}
   m}{\sqrt{\frac{q}{\sqrt{F}}}}-4\right)\right)}{15 \sqrt[4]{2} q^2 \sqrt{\frac{q}{\sqrt{F}}}},\\
 h^{(H)}(\phi) &=& -\frac{3}{2} a a_\Sigma \sec \phi ,\\
  V^{(H)}(\phi) &=& \frac{a_\Sigma \cos ^4\phi  \left(2 a
   \sec\phi +3 m\right)}{4 a^2}. 
\end{eqnarray}

\begin{figure}
    \centering
    \includegraphics[width=1\linewidth]{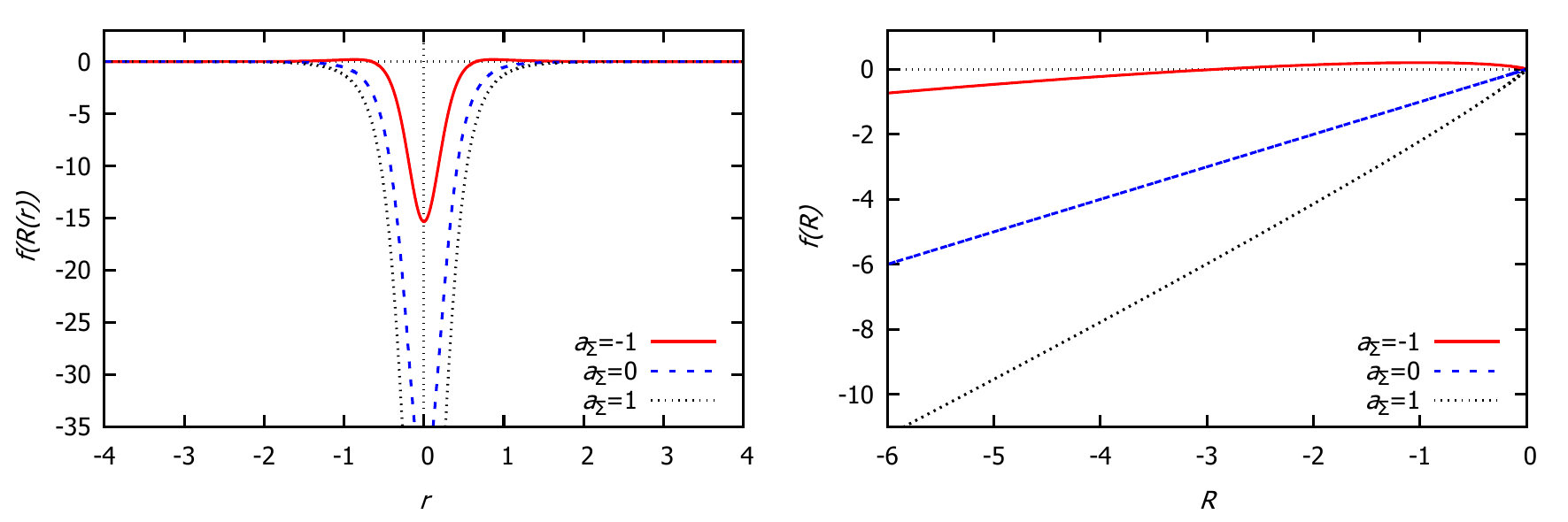}
    \caption{Behavior of the function $f(R)$ as a function of the radial coordinate (left panel) and as a function of the curvature scalar (right panel), fixing $a = m = 1$ and varying the values of the parameter $a_1$. In this case, the function $f(R(r))$ is symmetric under the transformation $r \to -r$, even when $a_\Sigma \neq 0$. This ensures that we obtain a single value of $f(R)$ for each value of the curvature scalar, thereby avoiding any multivalued behavior.}
    \label{fig:fR_fRr_model4}
\end{figure}

Considering the curvature scalar for the SV case, we can write the function $H(R(r))$ as:
\begin{equation}
    H(R(r))=-2 a^2 \left(\frac{15 a_\Sigma m}{4 \left(a^2+r^2\right)^2} -\frac{4 a_\Sigma}{3 \left(a^2+r^2\right)^{3/2}}\right).
\end{equation}
This function is symmetric under the transformation $r \to -r$. As in the previous cases, we cannot express the behavior of $f(R)$ analytically; therefore, we will analyze it graphically in Fig. \ref{fig:fR_fRr_model4}. From the behavior of $f(R(r))$, we explicitly observe the symmetry of the solution under the transformation $r \to -r$. From the parametric plot, we see that, as expected, for $a_\Sigma = 0$, we obtain a straight line, which corresponds to the case of GR.

We can generalize this model to $H_R =  a_\Sigma \Sigma(r)^n$. The issue with this generalization is that some functions exhibit divergences for certain values of $n$. Interestingly, these divergences cancel out when combining $f(R) + L(F(r)) + 2V(\phi (r))$, which is precisely the combination that appears in the action. The remaining terms of the action do not present divergences.

With this, we can prove that it is possible to find different types of matter sources for the SV metric by considering different $f(R)$ models.

\subsubsection{A brief comment about viability conditions and scalaron mass for case IV}
In this case, since $f_R(R(r)) = 1 + a_\Sigma \sqrt{r^2 + a^2}$, the condition $f_{R} > 0$ is satisfied for all regions if $a_\Sigma > 0$. If $a_\Sigma < 0$, this condition is only satisfied in the regions where $|r| < \sqrt{\frac{1}{a_\Sigma^2} - a^2}$. For the condition $f_{RR}>0$, we must write
\begin{equation}
    f_{RR}(R(r)) = -\frac{a_\Sigma r \left(a^2+r^2\right)^2}{2 a^2 \left(\sqrt{a^2+r^2}-3 m\right)}.
\end{equation}
In this case, the sign of $f_{RR}$ has the same dependence as in case III, simply replacing the coupling constant $a_2$ with $a_\Sigma$, so that the regions where $f_{RR} > 0$ are the same for both cases.

\begin{figure}[ht]
    \centering
    \includegraphics[width=0.5\linewidth]{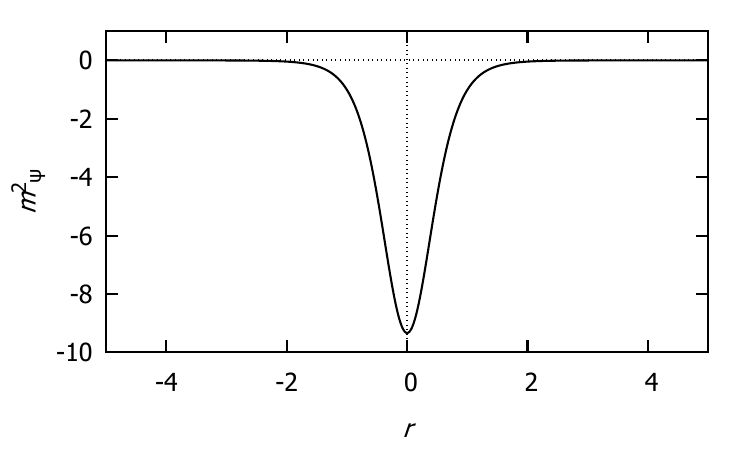}\includegraphics[width=0.5\linewidth]{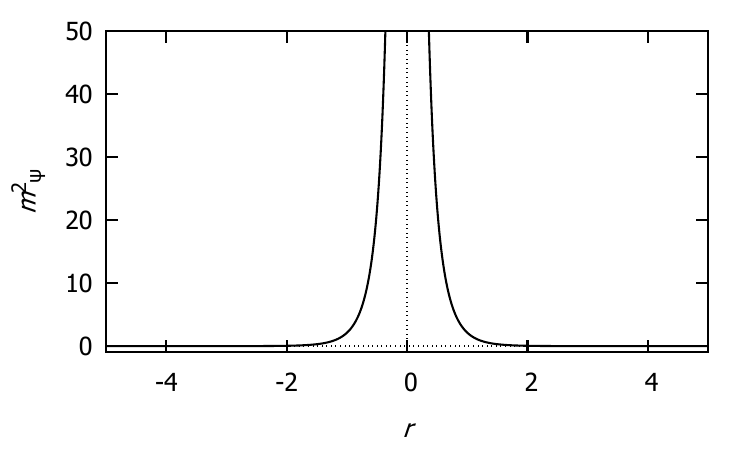}
    \caption{Scalaron mass $m_\psi^2$ as a function of the radial coordinate $r$ for $a_2 = 1$ (left) and $a_2 = -1$ (right), with $a = m = 1$. In this case, we have \( m_\psi^2 < 0 \) in center regions for \( a_\Sigma > 0 \), while for \( a_\Sigma < 0 \) we have \( m_\psi^2 > 0 \) in center regions, indicating local stability.}
    \label{fig:m2psi4}
\end{figure}

Finally, we can write the scalaron mass for the last case as
\begin{equation}\label{scalaron3}
 m_{\psi}^{2}(R(r)) =   \frac{2 a^2 \left(-3 m \left(8 a_\Sigma \sqrt{a^2+r^2}+5\right)+7 a^2 a_\Sigma+4 \sqrt{a^2+r^2}+7 a_\Sigma r^2\right)}{3 a_\Sigma \left(a^2+r^2\right)^3 \left(a_\Sigma
   \sqrt{a^2+r^2}+1\right)}.
\end{equation}
Observing Fig.~\ref{fig:m2psi4}, we see that, in more central regions, $m_\psi^2<0$ if $a_\Sigma>0$, 
whereas $m_\psi^2>0$ if $a_\Sigma<0$. However, from \eqref{scalaron3}, for $a_\Sigma>0$ and large $r$, the negative term grows as $r$ while the positive term grows as $r^2$; therefore, 
at large distances we have $m_\psi^2>0$ for $a_\Sigma>0$ and $m_\psi^2<0$ for $a_\Sigma<0$.

\section{Energy Conditions}\label{S:energy}

In the context of GR, it is well-known that BB geometries violate all energy conditions associated with the stress-energy tensor \cite{Simpson:2018tsi,Lobo:2020ffi}. This is not entirely surprising, as BB geometries interpolate between regular BH space-times and wormholes, both of which are known to violate the energy conditions \cite{Bronnikov:2018vbs}. Thus, it is interesting to investigate whether this violation persists in the context of modified theories of gravity, as in our case.

We will now analyze whether, in the context of our $f(R)$ models, the BB geometry satisfies the null energy condition (NEC), strong energy condition (SEC), weak energy condition (WEC) and dominant energy condition (DEC). These conditions are given by \cite{Visser:1995cc}:
\begin{eqnarray}
&&NEC_{1,2}=WEC_{1,2}=SEC_{1,2} 
\Longleftrightarrow \rho+p_{r,t}\geq 0,\label{Econd1} \\
&&SEC_3 \Longleftrightarrow\rho+p_r+2p_t\geq 0,\label{Econd2}\\
&&DEC_{1,2} \Longleftrightarrow \rho-|p_{r,t}|\geq 0 \Longleftrightarrow 
(\rho+p_{r,t}\geq 0) \hbox{ and } (\rho-p_{r,t}\geq 0),\label{Econd3}\\
&&DEC_3=WEC_3 \Longleftrightarrow\rho\geq 0,\label{Econd4}
\end{eqnarray}
where $\rho$, $p_r$, and $p_t$ are the energy density, radial pressure, and tangential pressure, respectively \footnote{Typically, we use the subscript $1$ ($2$) to denote cases where the energy density is combined solely with the radial (tangential) pressure. The subscript $3$ is employed either when considering the energy density alone or when combining all three functions.}. These functions are identified by the components of the stress-energy tensor as
\begin{equation}
    T^\mu_\nu = \text{diag}[\rho, -p_r,-p_t,-p_t].
\end{equation}
This relation is valid in regions where $A(r) > 0$. For regions where $A(r) < 0$, the signature of the metric changes, and we have:
\begin{equation}
    T^\mu_\nu = \text{diag}[-p_r, \rho,-p_t,-p_t].
\end{equation}

\subsection{Case I}
Let us now analyze the energy conditions for the case where $H(R) = a_R R^2$. For regions where $A(r) > 0$, we have:
\begin{eqnarray}
    \rho + p_r &=&\frac{2 a^2a \left(2m-\sqrt{a^2+r^2}\right)}{\left(a^2+r^2\right)^{5/2}}+ \frac{4 a^2 a_R \left(a^2-10 r^2\right) \left(2 a^2+18 m^2+2 r^2-13 m \sqrt{a^2+r^2}\right)}{\left(a^2+r^2\right)^5}, \\
    \rho + p_t &=&\frac{3 a^2
   m}{\left(a^2+r^2\right)^{5/2}}+4a^2 a_R\left[\frac{  a^2 \left(4 r^2-9 m^2\right)+45 m^2 r^2+4 r^4}{\left(a^2+r^2\right)^5}+\frac{3  m \left(a^2-9 r^2\right)}{\left(a^2+r^2\right)^{9/2}}\right],\\
\rho + p_r + 2p_t &=&\frac{2
   a^2 m}{\left(a^2+r^2\right)^{5/2}}+4 a^2 a_R\left[\frac{ r^2 \left(7 a^2+180 m^2\right)-5 a^2 \left(a^2+9 m^2\right)+12 r^4}{\left(a^2+r^2\right)^5}+\frac{ m \left(31 a^2-100 r^2\right)}{\left(a^2+r^2\right)^{9/2}}\right],\\
\rho - p_r &=&\frac{4 a^2 m}{\left(a^2+r^2\right)^{5/2}}
+4 a^2 a_R\left[\frac{ m \left(46 r^2-25 a^2\right)}{\left(a^2+r^2\right)^{9/2}}+\frac{5 a^4+a^2 \left(27 m^2+r^2\right)-90 m^2 r^2-4
   r^4}{\left(a^2+r^2\right)^5}\right],\\
    \rho - p_t &=& \frac{5 m a^2}{\left(a^2+r^2\right)^{5/2}}-\frac{2a^2a_0}{\left(a^2+r^2\right)^2}\nonumber\\
    &+&4 a^2 a_R\left[\frac{ m \left(203 r^2-41 a^2\right)}{\left(a^2+r^2\right)^{9/2}}+\frac{ 7 a^4-21 r^2
   \left(a^2+15 m^2\right)+54 a^2 m^2-28 r^4}{\left(a^2+r^2\right)^5}\right],\\
\rho &=& \frac{a^2 \left(4 m-\sqrt{a^2+r^2}\right)}{\left(a^2+r^2\right)^{5/2}}
-a^2 a_R\left[\frac{4  m \left(19 a^2-88 r^2\right)}{\left(a^2+r^2\right)^{9/2}}+\frac{ -14 a^4+a^2 \left(34 r^2-90
   m^2\right)+540 m^2 r^2+48 r^4}{\left(a^2+r^2\right)^5}\right].\nonumber\\
\end{eqnarray}
In regions where $A(r)<0$, we have
\begin{eqnarray}
    \rho + p_r &=& \frac{2 a^2 \left(\sqrt{a^2+r^2}-2 m\right)}{\left(a^2+r^2\right)^{5/2}}-\frac{4 a^2 a_R \left(a^2-10 r^2\right) \left(2 a^2+18 m^2+2 r^2-13 m \sqrt{a^2+r^2}\right)}{\left(a^2+r^2\right)^5},\\
    \rho + p_t &=& \frac{a^2 \left(2 \sqrt{a^2+r^2}-m\right)}{\left(a^2+r^2\right)^{5/2}}+4a^2 a_R\left[\frac{ m\left(16 a^2 -157 r^2\right)}{\left(a^2+r^2\right)^{9/2}}+\frac{ \left(2 a^4+a^2 \left(22 r^2-27
   m^2\right)+225 m^2 r^2+24 r^4\right)}{\left(a^2+r^2\right)^5}\right],\nonumber\\\\
\rho + p_r + 2p_t &=& \frac{4 a^2}{\left(a^2+r^2\right)^2}-\frac{6 a^2 m}{\left(a^2+r^2\right)^{5/2}}\nonumber\\
&+& 4 a^2 a_R\left[\frac{ m \left(360 r^2-57 a^2\right)}{\left(a^2+r^2\right)^{9/2}}+\frac{ \left(9 a^4+a^2
   \left(81 m^2-43 r^2\right)-540 m^2 r^2-52 r^4\right)}{\left(a^2+r^2\right)^5}\right],\\ 
\rho - p_r &=& \frac{4 a^2 m}{\left(a^2+r^2\right)^{5/2}}
+4 a^2 a_R\left[\frac{ m \left(46 r^2-25 a^2\right)}{\left(a^2+r^2\right)^{9/2}}+\frac{ \left(5 a^4+a^2 \left(27 m^2+r^2\right)-90 m^2 r^2-4
   r^4\right)}{\left(a^2+r^2\right)^5}\right],\\  
\rho - p_t &=&\frac{a^2 m}{\left(a^2+r^2\right)^{5/2}}+4 a^2 a_R\left[\frac{ m \left(73 r^2-28 a^2\right)}{\left(a^2+r^2\right)^{9/2}}+\frac{ \left(5 a^4-3 r^2 \left(a^2+45 m^2\right)+36 a^2 m^2-8
   r^4\right)}{\left(a^2+r^2\right)^5}\right],\\
\rho &=& \frac{a^2}{\left(a^2+r^2\right)^2}
-2 a^2 a_R\left[\frac{12  m \left(a^2+7 r^2\right)}{\left(a^2+r^2\right)^{9/2}}-\frac{ \left(3 a^4+a^2 \left(9 m^2+19 r^2\right)+90 m^2 r^2+16
   r^4\right)}{\left(a^2+r^2\right)^5}\right].
\end{eqnarray}
The terms involving $a_R$ represent the corrections from the $f(R)$ theory to the energy conditions of this model. However, the analytical expressions do not clearly illustrate how the $f(R)$ theory corrections modify the energy conditions. 

\begin{figure}
    \centering
    \includegraphics[width=1\linewidth]{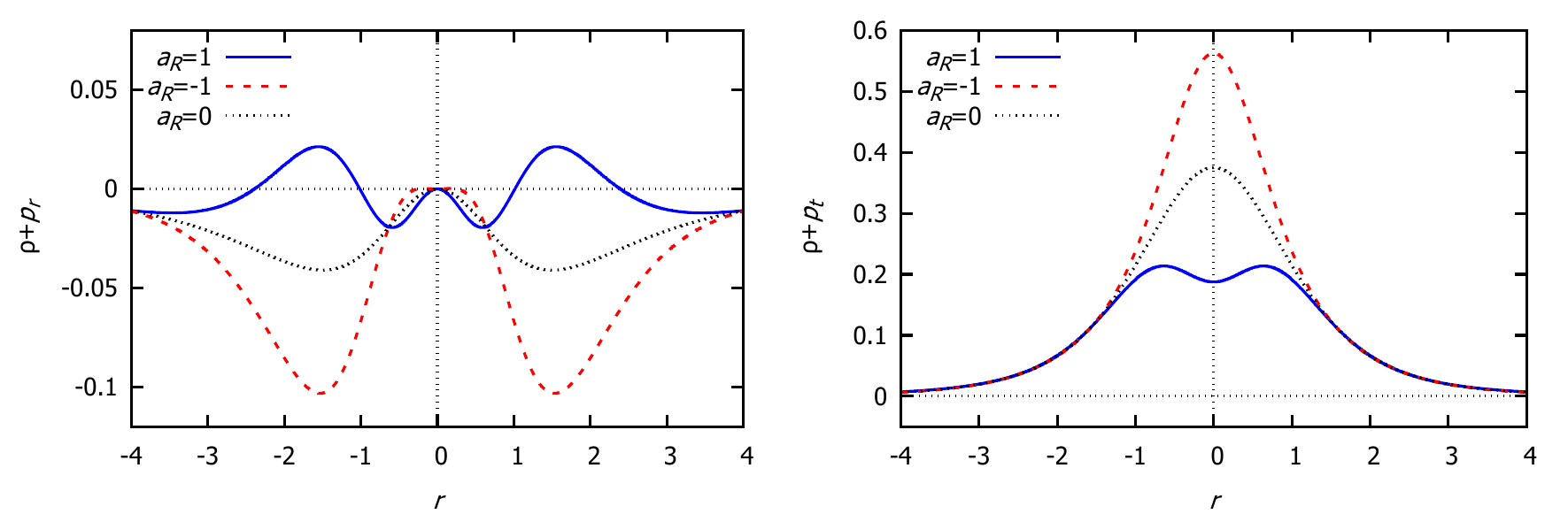}
    \includegraphics[width=1\linewidth]{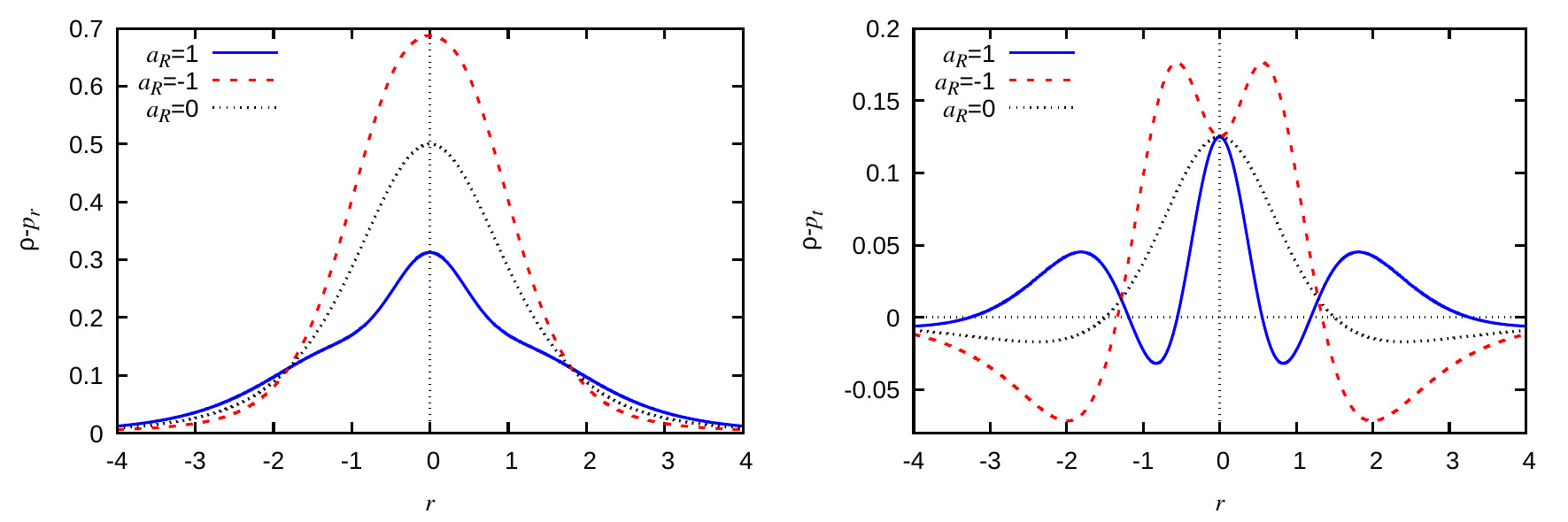}
      \includegraphics[width=1\linewidth]{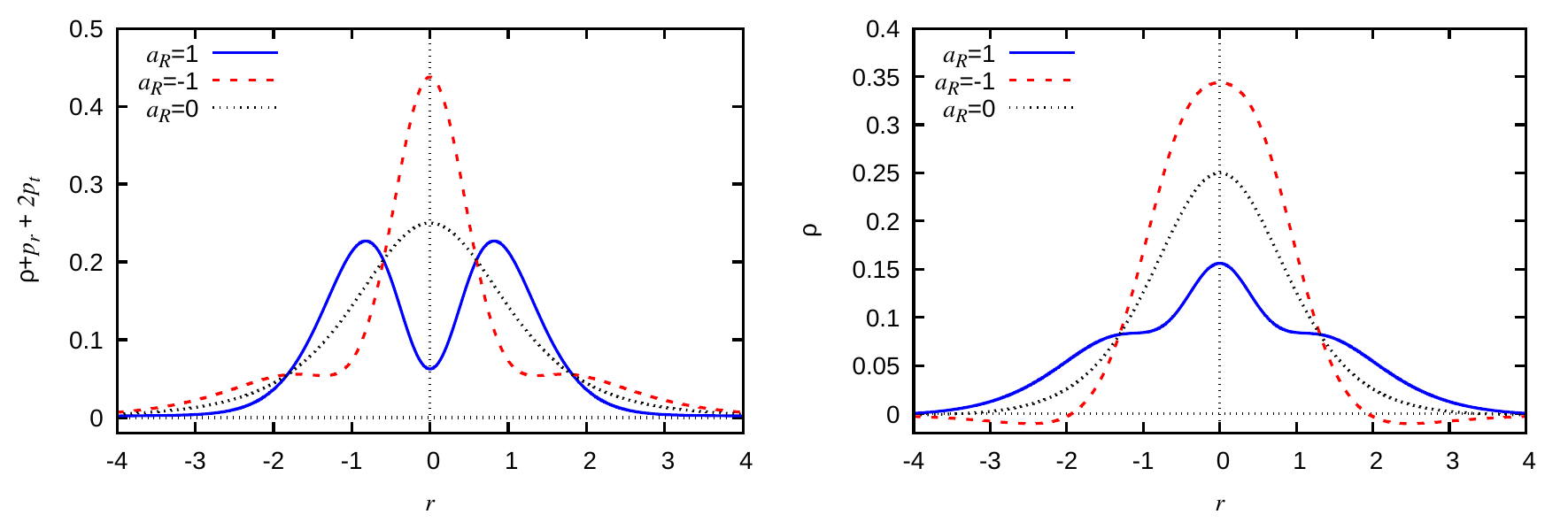}
    \caption{Combinations of the stress-energy tensor components for the $H(R) = a_R R^2$ model as a function of the radial coordinate, with $a = 2$ and $m = 1$, for different values of $a_R$. For the chosen set of parameters, the horizon is located at the same point as the throat, at $r=0$. Depending on the value of the constant $a_R$, the regions where these combinations are positive or negative may change.}
    \label{fig:EC_model1}
\end{figure}

In Fig. \ref{fig:EC_model1}, we compare how the $f(R)$ theory modifies the energy conditions relative to GR. We observe that for $a_R = -1$, the combination $\rho + p_r$ is always negative, with a magnitude even greater than in the case of GR. However, for $a_R = 1$, there are certain regions where $\rho + p_r$ takes positive values, indicating that the $f(R)$ theory relaxes the violation of the null energy condition. From the behavior of $\rho$ and $\rho - p_t$, we observe that depending on the values of the chosen parameters, in some regions where the inequalities were satisfied in the case of GR, they become violated, and in some regions where they were violated, they become satisfied. Thus, the contributions from $H(R)$, in this case, the model $H(R) = a_R R^2$, can result in the energy conditions not being always violated, as was the case in GR.

\subsection{Case II}

In this case, where $H_R = a_1r$, we have, outside any possible horizon, $A(r)>0$, that the following combinations for the functions $\rho$, $p_r$, and $p_t$ are

\begin{eqnarray}
\rho + p_r &=& -\frac{2 a^2 \left(1 + a_1 r\right) \left( \sqrt{a^2 + r^2}-2 m \right)}{\left(a^2 + r^2\right)^{5/2}},\\
\rho + p_t &=& \frac{3 a^2  m}{\left(a^2+r^2\right)^{5/2}}+\frac{a_1 r \left(6 a^2 m-\left(a^2+r^2\right)^{3/2}+3 m r^2\right)}{\left(a^2+r^2\right)^{5/2}},\\
\rho + p_r + 2p_t &=& \frac{2 a^2  m}{\left(a^2+r^2\right)^{5/2}}+a_1\left(\frac{r}{a^2+r^2}+\frac{2  m r \left(4 a^4+5 a^2 r^2+2
   r^4\right)}{a^2 \left(a^2+r^2\right)^{5/2}}+\frac{2  \tan ^{-1}\left(\frac{\sqrt{a^2+r^2}-r}{a}\right)}{a}\right),\\
\rho - p_r &=& \frac{4 a^2  m}{\left(a^2+r^2\right)^{5/2}}+\frac{a_1}{a^2} \left(\frac{4 m r \left(a^4+a^2
   r^2-r^4\right)}{\left(a^2+r^2\right)^{5/2}}-\frac{3 a^2 r}{a^2+r^2}
   +2 a \tan ^{-1}\left(\frac{r-\sqrt{a^2+r^2}}{a}\right)\right),\\
\rho - p_t &=&  \frac{5 ma^2 }{\left(a^2+r^2\right)^{5/2}}-\frac{2a^2 a_0}{\left(a^2+r^2\right)^2}+\frac{a_1}{a^2} \left(\frac{m r \left(2 a^4-7 a^2 r^2-4 r^4\right)}{\left(a^2+r^2\right)^{5/2}}-\frac{2 a^2 r \left(2 a^2+r^2\right)}{\left(a^2+r^2\right)^2}\right.\nonumber\\
&+&\left.2 a \tan
   ^{-1}\left(\frac{r-\sqrt{a^2+r^2}}{a}\right)\right),\\
\rho &=& \frac{a^2  \left(4m-\sqrt{a^2+r^2}\right)}{\left(a^2+r^2\right)^{5/2}}+\frac{a_1}{2 a^2} \left(\frac{4 m r \left(2 a^4-a^2 r^2-r^4\right)}{\left(a^2+r^2\right)^{5/2}}-\frac{r \left(5 a^4+3 a^2
   r^2\right)}{\left(a^2+r^2\right)^2}\right.\nonumber\\
&+&\left.2 a \tan ^{-1}\left(\frac{r-\sqrt{a^2+r^2}}{a}\right)\right).
\end{eqnarray}

Whereas, for $A < 0$, we have:
\begin{eqnarray}
    \rho + p_r &=& \frac{2 a^2 \left(1 + a_1 r\right) \left( \sqrt{a^2 + r^2}-2 m \right)}{\left(a^2 + r^2\right)^{5/2}},\\
    \rho + p_t &=& \frac{
  a^2  \left(2 \sqrt{a^2 + r^2}-m\right)
  }{(a^2 + r^2)^{5/2}}+\frac{
  a_1 \left[r^3 \left(3 m - \sqrt{a^2 + r^2}\right) + 
    a^2 r \left(2 m + \sqrt{a^2 + r^2}\right)\right]
  }{(a^2 + r^2)^{5/2}} 
,\\
  \rho + p_r + 2p_t &=& \frac{
   \left(-6 a^4 m + 4 a^4 \sqrt{a^2 + r^2}\right)
  }{a^2 (a^2 + r^2)^{5/2}} 
+ \frac{
  a_1}{a^2 (a^2 + r^2)^{5/2}} \left[4 m r^5 + 5 a^4 r \sqrt{a^2 + r^2}\right.\nonumber\\
  &+&\left. 
    a^2 r^3 \left(10 m + \sqrt{a^2 + r^2}\right) + 
    2 a (a^2 + r^2)^{5/2} \arctan\left(\frac{-r + \sqrt{a^2 + r^2}}{a}\right)\right],\\
\rho - p_r &=&  \frac{4 a^2  m}{\left(a^2 + r^2\right)^{5/2}} + \frac{a_1}{a^2 \left(a^2 + r^2\right)^{5/2}} \left[4 a^4 m r - 4 m r^5 - 3 a^4 r \sqrt{a^2 + r^2}\right.\nonumber\\
&-& \left.a^2 r^3 \left(4 m + 3 \sqrt{a^2 + r^2}\right) + 2a(a^2 + r^2)^{5/2} \arctan\left(\frac{\sqrt{r^2 + a^2} - r}{a}\right)\right],\\
\rho - p_t &=& \frac{
  a^2  m
  }{(a^2 + r^2)^{5/2}} 
+ \frac{
  a_1}{a^2 (a^2 + r^2)^{5/2}} \left[-4 m r^5 - 2 a^4 r \left(m + \sqrt{a^2 + r^2}\right)\right.\nonumber\\
  &-&\left. 
    a^2 r^3 \left(7 m + 2 \sqrt{a^2 + r^2}\right) + 
    2 a (a^2 + r^2)^{5/2} \arctan\left(\frac{r - \sqrt{a^2 + r^2}}{a}\right)\right],\\
\rho &=& \frac{a^2 }{\left(a^2+r^2\right)^2}-\frac{a_1}{2 a^2} \left(\frac{4 m r^3}{\left(a^2+r^2\right)^{3/2}}+\frac{r a^2\left(a^2+3 
   r^2\right)}{\left(a^2+r^2\right)^2}-2 a \tan ^{-1}\left(\frac{r-\sqrt{a^2+r^2}}{a}\right)\right).
\end{eqnarray}
The terms where $a_1$ appears represent the corrections from the $f(R)$ theory to the energy conditions. We can extract some information from the analytical expressions of $\rho + p_r$. If $a_1 > 0$ and we consider the region where $r > 0$, the null energy condition will be violated in the same way as in GR, whereas for $r < 0$, the $NEC_1$ inequality will be satisfied if $r < -1 / a_1$. For $a_1 < 0$, the effect is reversed. The inequality $NEC_1$ will be satisfied if $r > -1 / a_1$, and is violated in the other regions. We will analyze the energy conditions graphically to obtain more insights about the energy conditions.
\begin{figure}[!htb]
    \centering
    \includegraphics[width=1\linewidth]{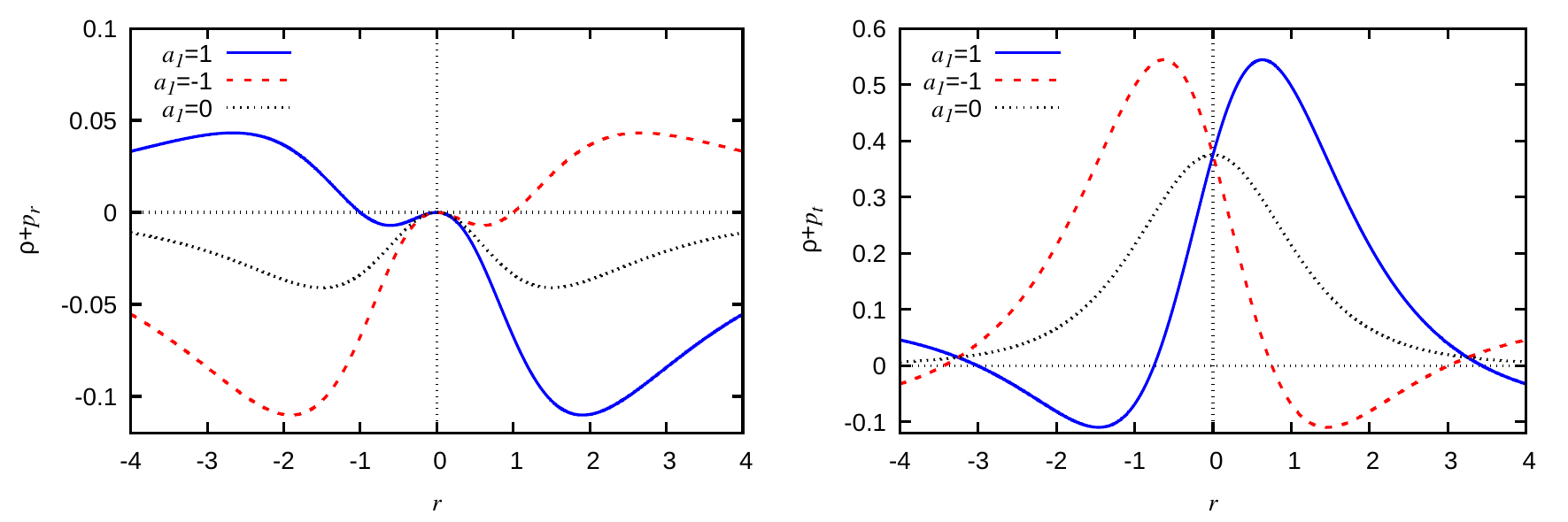}
    \includegraphics[width=1\linewidth]{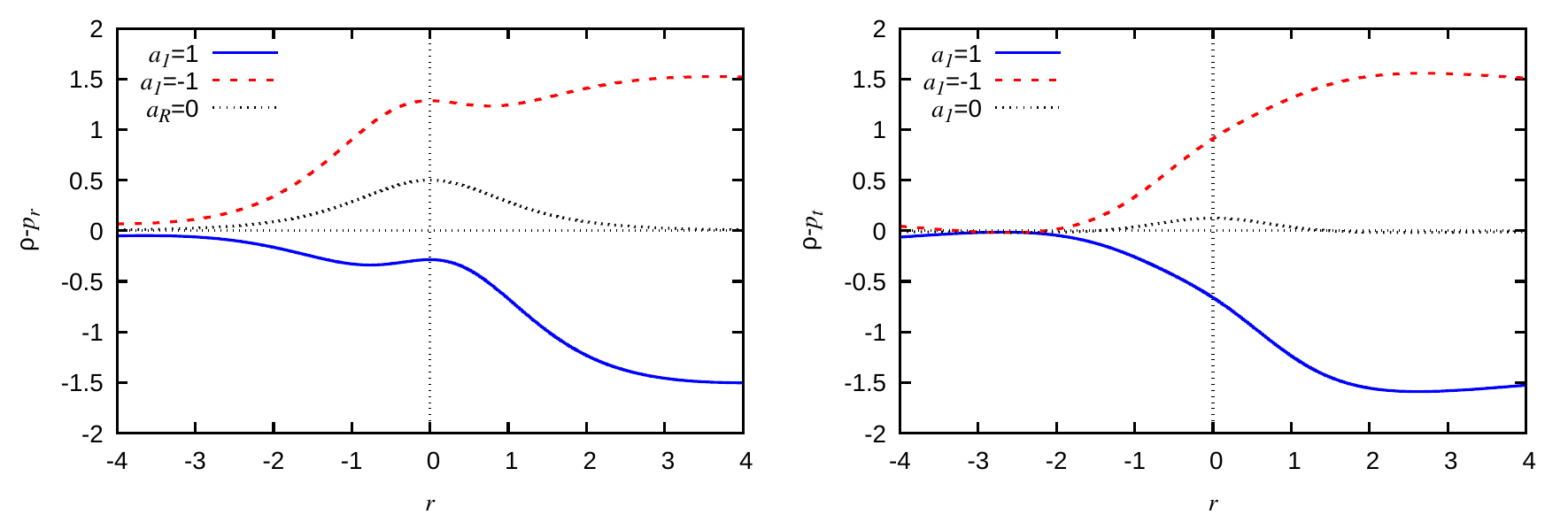}
      \includegraphics[width=1\linewidth]{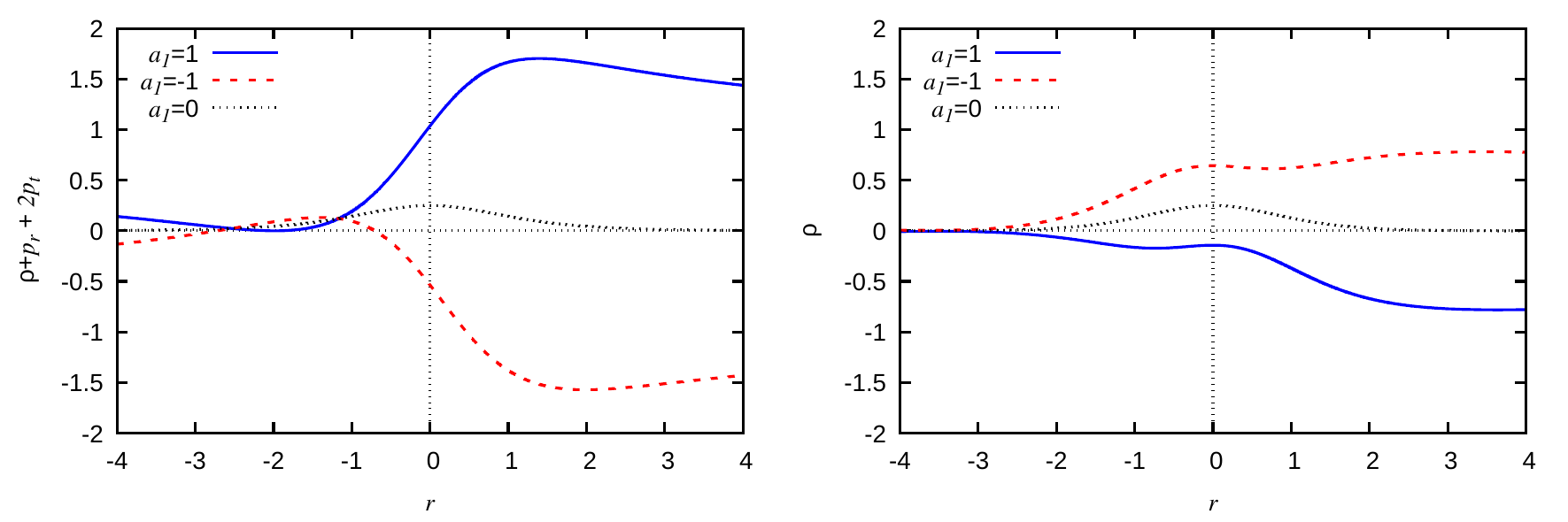}
    \caption{Combination of the stress-energy tensor components for the model $H_R =  a_1 r$, as a function of the radial coordinate, for $m = 1$, $a = 2$, and different values of $a_1$. For the chosen set of parameters, the horizon is located at the same point as the throat, at $r=0$. Depending on the value of the constant $a_1$, the regions where these combinations are positive or negative may change. Since the components of the stress-energy tensor depend on the functions associated with the $f(R)$ theory, and these functions are asymmetric under the transformation $r \to -r$, the combinations of the stress-energy tensor components are likewise not symmetric.}
    \label{fig:EC_model2}
\end{figure}

In Fig. \ref{fig:EC_model2}, we present the graphical representation of the energy density and its combinations with the pressures as functions of the coordinate $r$ for different values of the parameter $a_1$. Analyzing the figure, we see that the conditions $WEC_3 \geq 0 $ are always satisfied for $a_1 = 0$ and $a_1 = - 1$, while is violated throughout the entire space-time for $a_1 = 1$. On the other hand, it is possible to observe that the $NEC_{1,2}$ condition is satisfied only in certain regions of space-time, both for $a_1 = -1$ and $a_1 = 1$. Regarding condition $SEC_3$, we see that it is satisfied for all values of $r$ for $a_1 = 1$ and $a_1 = 0$. Regarding condition $DEC_{1,2}$, although $\rho - p_{r,t} \geq 0$ for $a_1 = - 1$ and $a_1 = 0$ for all $r$, we find that this condition is satisfied only in certain regions, since $\rho + p_{r,t} \geq 0$ only in certain regions.

Thus, we conclude that, depending on the chosen values, this form of the $f(R)$ theory contributes to the satisfaction of the energy conditions, at least in certain regions of space-time.

\subsection{Case III}
Let us now consider the corrections to the energy conditions for the model $H_R = a_2 r^2$. In regions where $A(r) > 0 $, we have
\begin{eqnarray}
    \rho + p_r &=& -\frac{2 \left[a^4 a_2 + a_2 r^4 + a^2 \left(1 + 3 a_2 r^2\right)\right] \left[a^2 + r^2 + 4 m \left(m - \sqrt{a^2 + r^2}\right)\right]}{\left(a^2 + r^2\right)^{5/2} \left(-2 m + \sqrt{a^2 + r^2}\right)},\\
    \rho + p_t &=& \frac{-2 a_2 r^4 \left(-3 m + \sqrt{a^2 + r^2}\right) + a^2 \left[3 m + a_2 r^2 \left(9 m - 2 \sqrt{a^2 + r^2}\right)\right]}{\left(a^2 + r^2\right)^{5/2}},\\
    \rho + p_r + 2p_t &=& -\frac{2 a^2  m}{\left(a^2+r^2\right)^{5/2}}-a_2\left[\frac{  \left(2a^2+3 r^2\right)}{a^2+r^2}-\frac{ m \left(4 a^4+5 a^2 r^2+2 r^4\right)}{\left(a^2+r^2\right)^{5/2}}\right],\\
    \rho - p_r &=& -\frac{4 a^2  m}{\left(a^2+r^2\right)^{5/2}}+2a_2\left[\frac{ \left(2 a^2+5 r^2\right)}{a^2+r^2}-\frac{2 m \left(2 a^4+7 a^2 r^2+4 r^4\right)}{\left(a^2+r^2\right)^{5/2}}\right],\\
    \rho - p_t &=& \frac{a^2  \left(2 \sqrt{a^2+r^2}-5m\right)}{\left(a^2+r^2\right)^{5/2}}-a_2\left[\frac{m \left(a^2+2 r^2\right) \left(12 a^2+7 r^2\right)}{\left(a^2+r^2\right)^{5/2}}-\frac{2  \left(3 a^4+9
   a^2 r^2+5 r^4\right)}{\left(a^2+r^2\right)^2}\right], \\
    \rho &=& \frac{a^2  \left(4 m - \sqrt{a^2 + r^2}\right)}{\left(a^2 + r^2\right)^{5/2}} + \frac{a_2}{\left(a^2 + r^2\right)^{5/2}} \left[2 r^4 \left(5 m - 3 \sqrt{a^2 + r^2}\right)\right.\nonumber\\
    &-&\left. (3 a^4-10a^2r^2) \left(\sqrt{a^2 + r^2}-2 m \right) \right].\nonumber\\
\end{eqnarray}

\begin{figure}
    \centering
    \includegraphics[width=1\linewidth]{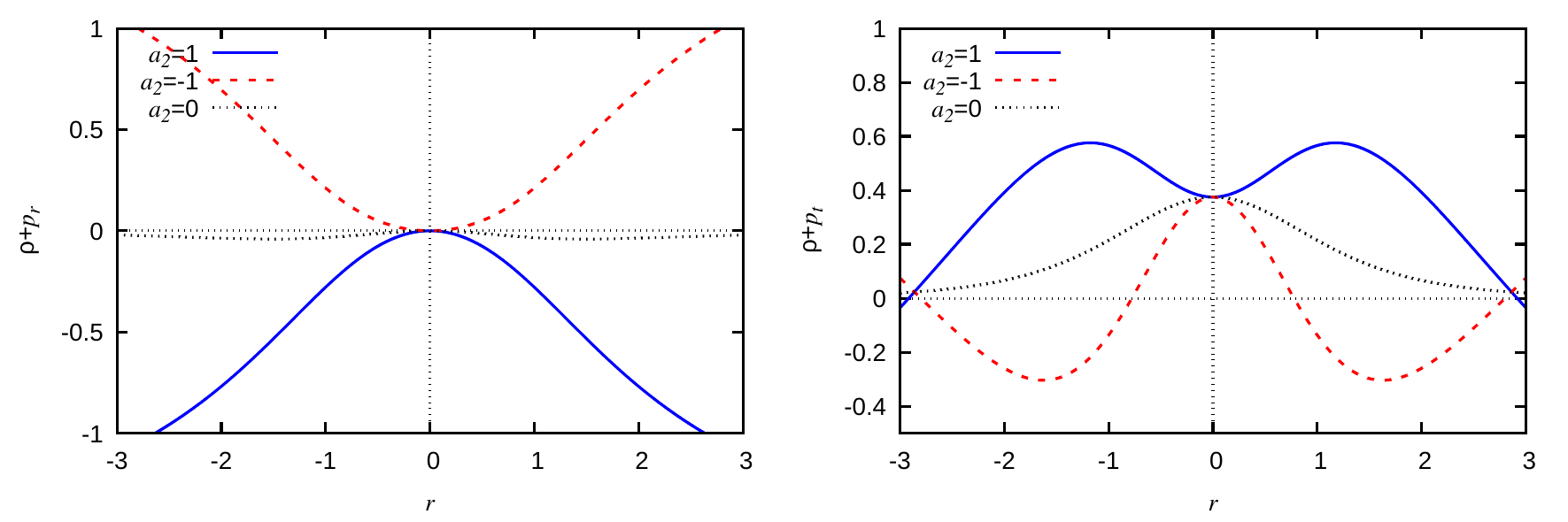}
    \includegraphics[width=1\linewidth]{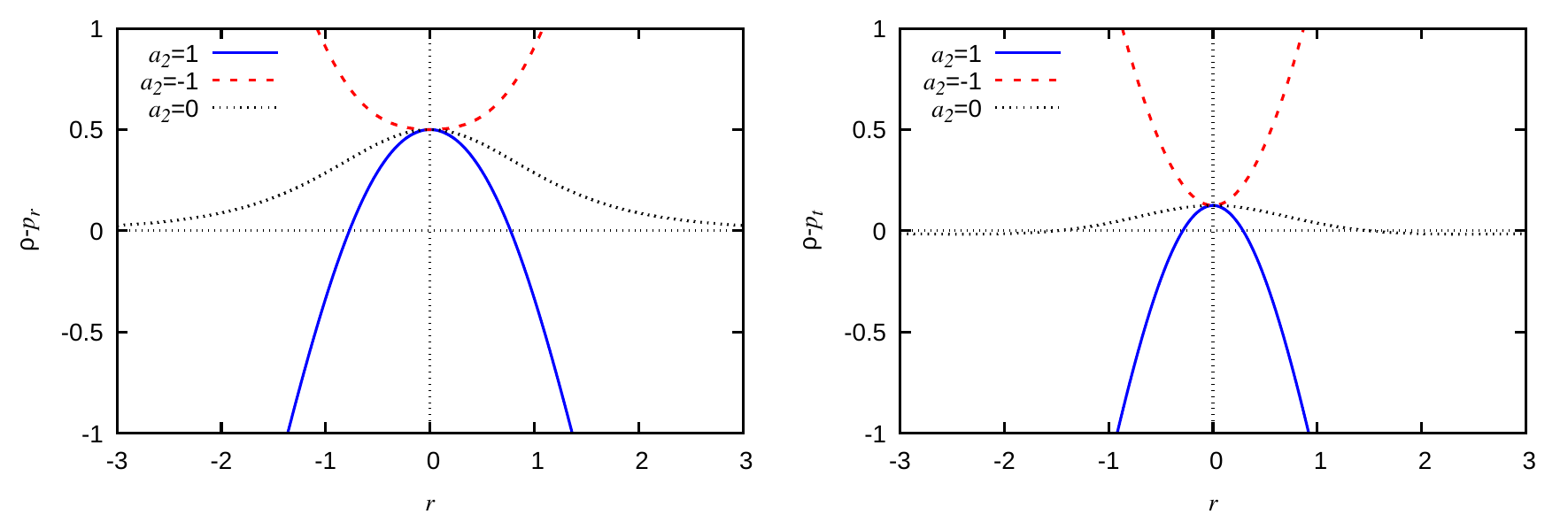}
      \includegraphics[width=1\linewidth]{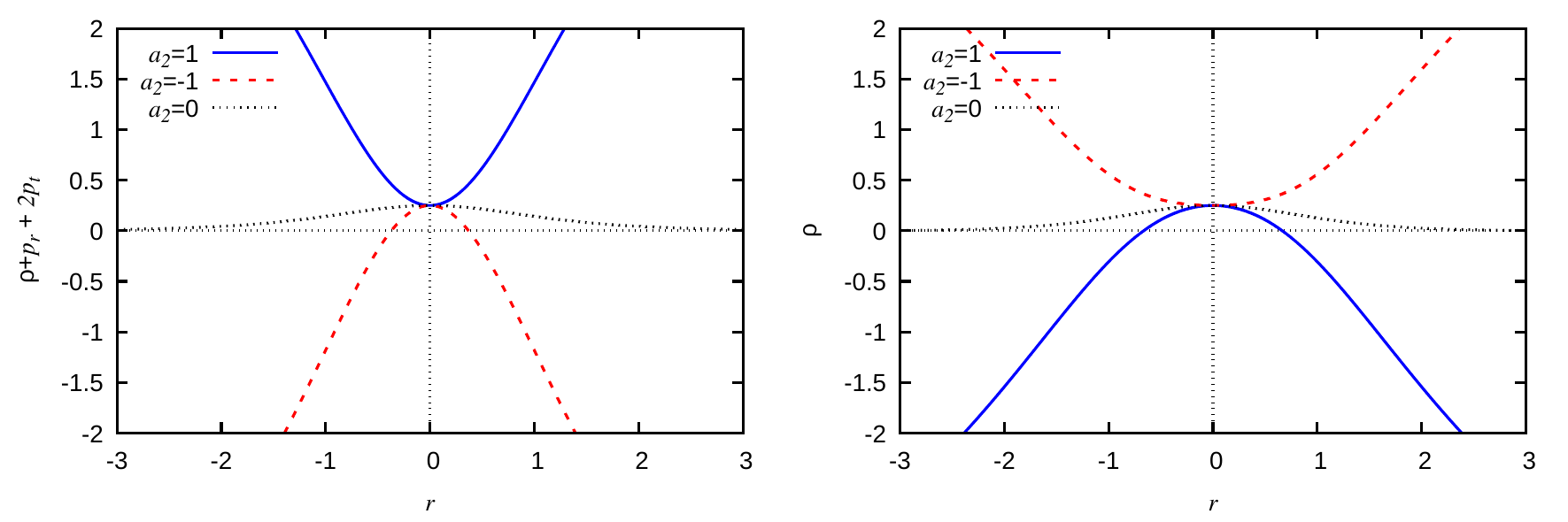}
    \caption{Combination of the stress-energy tensor components for the model $H_R = a_2 r^2$, as a function of the radial coordinate, for $m = 1$, $a = 2$, and different values of $a_2$. For the chosen set of parameters, the horizon is located at the same point as the throat, at $r=0$. Depending on the value of the constant $a_2$, the regions where these combinations are positive or negative may change.}
    \label{fig:EC_model3}
\end{figure}

In regions where $A(r) < 0$, we have
\begin{eqnarray}
 \rho + p_r &=& \frac{2 \left[a^4 a_2 + a_2 r^4 + a^2 \left(1 + 3 a_2 r^2\right)\right] \left[a^2 + r^2 + 4 m \left(m - \sqrt{a^2 + r^2}\right)\right]}{\left(a^2 + r^2\right)^{5/2} \left(-2 m + \sqrt{a^2 + r^2}\right)},\\
 \rho + p_t &=& \frac{
  a^2  \left(2 \sqrt{a^2 + r^2}-m \right)
  }{(a^2 + r^2)^{5/2}} 
+ a_2 \left[\frac{ m \left(-4 a^4-3 a^2 r^2+2 r^4\right)}{\left(a^2+r^2\right)^{5/2}}+\frac{2  \left(a^4+2 a^2 r^2\right)}{\left(a^2+r^2\right)^2}\right],\\
\rho + p_r + 2p_t &=& \frac{2 a^2  \left(2 \sqrt{a^2+r^2}-3 m\right)}{\left(a^2+r^2\right)^{5/2}}-2 a_2\left[\frac{ m \left(8 a^4+17 a^2 r^2+6 r^4\right)}{\left(a^2+r^2\right)^{5/2}}+\frac{ \left(4 a^4+11 a^2 r^2+5
   r^4\right)}{\left(a^2+r^2\right)^2}\right],\\
 \rho -p_r &=& \frac{4 a^2  m}{\left(a^2+r^2\right)^{5/2}}+2 a_2 \left(\frac{3 a^2}{a^2+r^2}-5\right)+\frac{4 a_2 m \left(2 a^4+7 a^2 r^2+4 r^4\right)}{\left(a^2+r^2\right)^{5/2}},\\   
\rho - p_t &=& \frac{a^2  m}{\left(a^2+r^2\right)^{5/2}}+4 a_2 \left(\frac{a^2}{a^2+r^2}-2\right)+\frac{a_2 m \left(8 a^4+19 a^2 r^2+10 r^4\right)}{\left(a^2+r^2\right)^{5/2}},\\
\rho &=& \frac{a^2 }{\left(a^2+r^2\right)^2}+\frac{2 a_2 m \left(a^2+3 r^2\right)}{\left(a^2+r^2\right)^{3/2}}-\frac{a_2 \left(a^2+2 r^2\right)^2}{\left(a^2+r^2\right)^2}.
\end{eqnarray}
The terms involving $a_2$ represent the corrections from the $f(R)$ theory to the energy conditions. The analytical expressions are unclear; therefore, we will prioritize analyzing them graphically.

In Fig. \ref{fig:EC_model3}, we show how the components of the energy-momentum tensor are modified by the presence of the $f(R)$ theory. Since our $f(R)$ theory model is symmetric under the transformation $r \to -r$, the energy conditions are also symmetric, allowing us to focus solely on the behavior in the $r > 0$ region. For the case $a_2 = 1$, as in GR, the condition $NEC_1$ is always violated. Disregarding the $SEC_3$ condition, the choice $a_2 = 1$ seems to increase the regions where the other conditions are violated compared to GR. For $a_2 = -1$, the $NEC_2$ condition is violated in more regions than in GR; however, this violation does not occur throughout the entire space-time. The $NEC_1$ condition is now satisfied in some regions. We can observe that there are regions where all combinations are positive. This means that, at least in some regions of space-time, all energy conditions are satisfied.
\subsection{Case IV}
Let us now analyze the energy conditions for the final model, $H_R =  a_\Sigma \Sigma$. In the regions where $A > 0$, we have
\begin{eqnarray}
    \rho + p_r &=& \frac{a^2  \left(4 m-2 \sqrt{a^2+r^2}\right)}{\left(a^2+r^2\right)^{5/2}}-\frac{3 a^2 a_\Sigma  \left(-2 m \sqrt{a^2+r^2}+a^2+r^2\right)}{\left(a^2+r^2\right)^{5/2}},\\
    \rho + p_t &=& \frac{3 a^2  m}{\left(a^2+r^2\right)^{5/2}}+a_\Sigma \left(\frac{3 m}{a^2+r^2}-\frac{r^2}{\left(a^2+r^2\right)^{3/2}}\right),\\
    \rho + p_r + 2p_t &=& \frac{2 a^2  m}{\left(a^2+r^2\right)^{5/2}}+\frac{a_\Sigma  \left(10 a^4+a^2 \left(22 r^2-9 m \sqrt{a^2+r^2}\right)+12 r^4\right)}{6 \left(a^2+r^2\right)^{5/2}},\\
    \rho - p_r &=& \frac{4 a^2  m}{\left(a^2+r^2\right)^{5/2}}+\frac{3 a_\Sigma  m \left(5 a^2+4 r^2\right)}{2 \left(a^2+r^2\right)^2}-\frac{a_\Sigma \left(5 a^2+12 r^2\right)}{3 \left(a^2+r^2\right)^{3/2}},\\
    \rho - p_t &=&  \frac{a^2  \left(5 m-2 \sqrt{a^2+r^2}\right)}{\left(a^2+r^2\right)^{5/2}}+\frac{3 a_\Sigma m \left(7 a^2+2 r^2\right)}{2 \left(a^2+r^2\right)^2}-\frac{a_\Sigma \left(14 a^2+9 r^2\right)}{3 \left(a^2+r^2\right)^{3/2}},\\
    \rho &=&  -\frac{a^2  \left(\sqrt{a^2+r^2}-4 m\right)}{\left(a^2+r^2\right)^{5/2}}+\frac{3 a_\Sigma m \left(9 a^2+4 r^2\right)}{4 \left(a^2+r^2\right)^2}-\frac{a_\Sigma  \left(7 a^2+6 r^2\right)}{3 \left(a^2+r^2\right)^{3/2}}.
\end{eqnarray}

\begin{figure}
    \centering
    \includegraphics[width=1\linewidth]{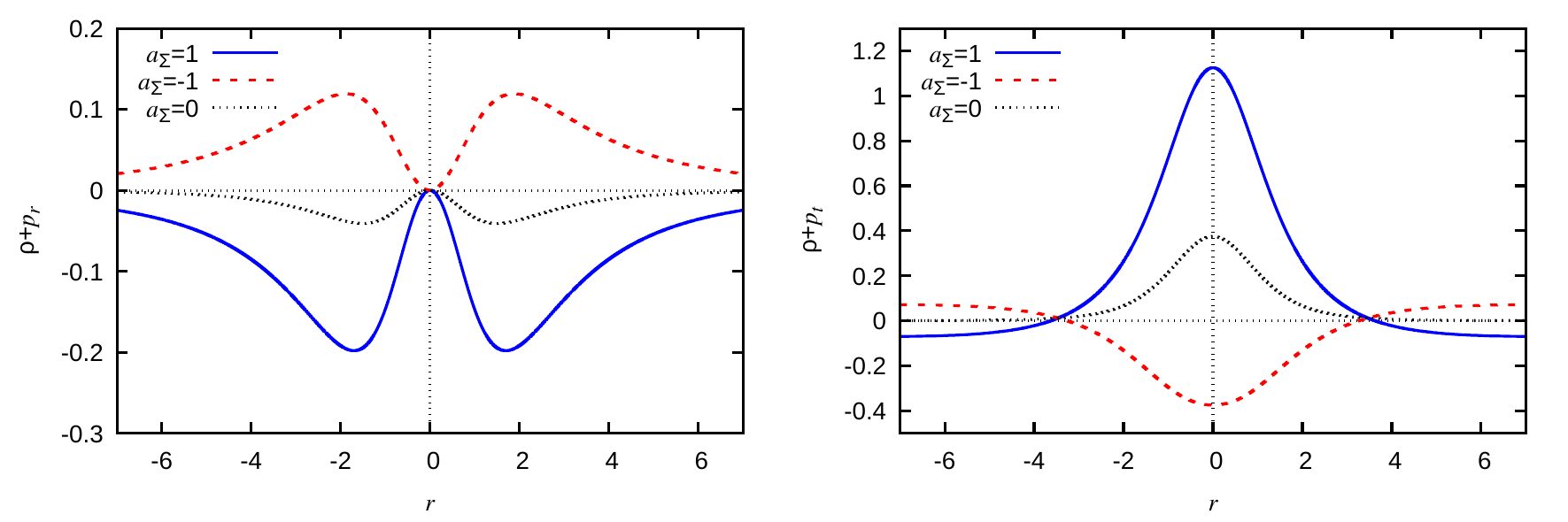}
    \includegraphics[width=1\linewidth]{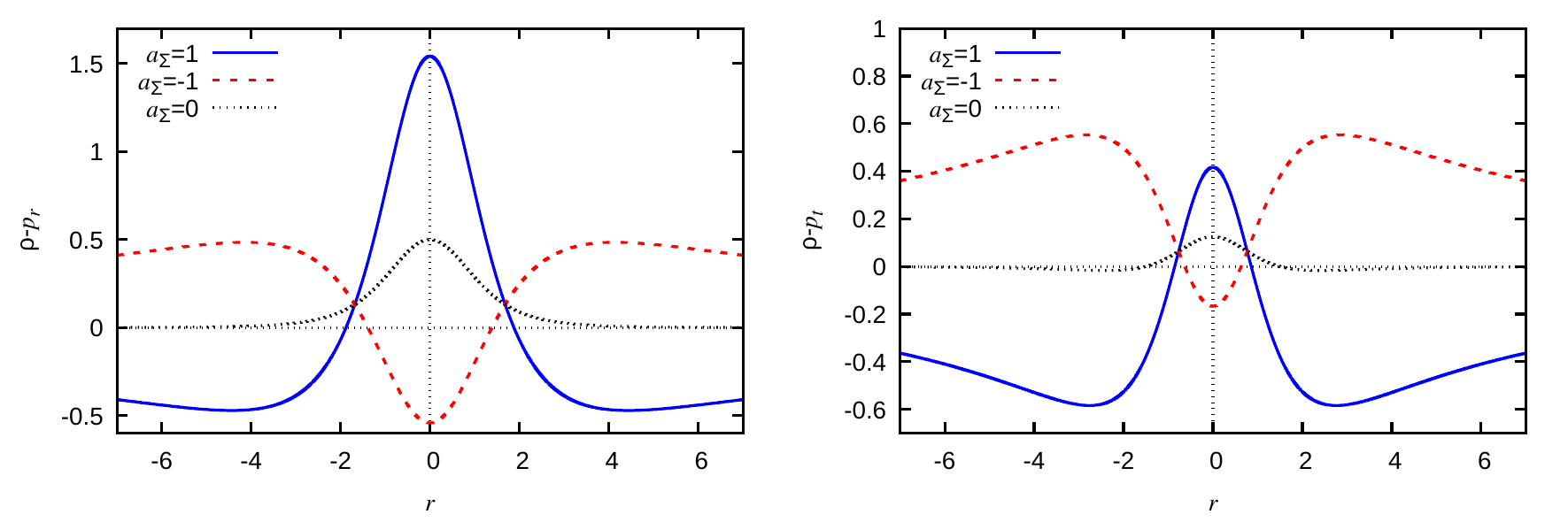}
      \includegraphics[width=1\linewidth]{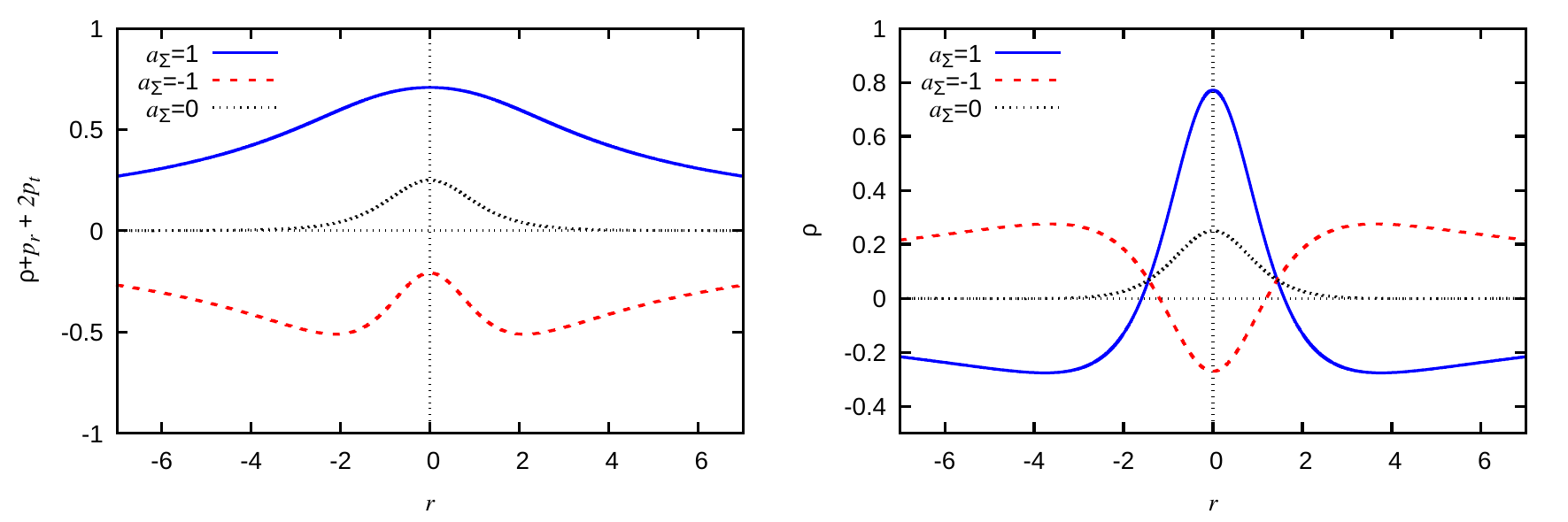}
    \caption{Combination of the stress-energy tensor components for the model $H_R =  a_\Sigma \Sigma $, as a function of the radial coordinate, for $m = 1$, $a = 2$, and different values of $a_\Sigma$. For the chosen set of parameters, the horizon is located at the same point as the throat, at $r=0$. Depending on the value of the constant $a_\Sigma$, the regions where these combinations are positive or negative may change.}
    \label{fig:EC_model4}
\end{figure}

In regions where $A<0$, we have
\begin{eqnarray}
    \rho + p_r &=& \frac{2 a^2  \left(\sqrt{a^2+r^2}-2 m\right)}{\left(a^2+r^2\right)^{5/2}}+\frac{3 a^2 a_\Sigma \left(-2 m \sqrt{a^2+r^2}+a^2+r^2\right)}{\left(a^2+r^2\right)^{5/2}},\\
\rho + p_t &=&\frac{a^2  \left(2 \sqrt{a^2+r^2}-m\right)}{\left(a^2+r^2\right)^{5/2}}+\frac{3 a_\Sigma  m \left(r^2-a^2\right)}{\left(a^2+r^2\right)^2}+\frac{a_\Sigma  \left(3
   a^2-r^2\right)}{\left(a^2+r^2\right)^{3/2}},\\
\rho + p_r + 2p_t &=& \frac{2 a^2  \left(2 \sqrt{a^2+r^2}-3 m\right)}{\left(a^2+r^2\right)^{5/2}}-\frac{27 a^2 a_\Sigma m}{2 \left(a^2+r^2\right)^2}+\frac{a_\Sigma \left(23 a^2+6 r^2\right)}{3
   \left(a^2+r^2\right)^{3/2}},\\
 \rho - p_r &=& \frac{4 a^2  m}{\left(a^2+r^2\right)^{5/2}}+\frac{3 a_\Sigma m \left(5 a^2+4 r^2\right)}{2 \left(a^2+r^2\right)^2}-\frac{a_\Sigma \left(5 a^2+12 r^2\right)}{3 \left(a^2+r^2\right)^{3/2}},\\
\rho - p_t &=& \frac{a^2  m}{\left(a^2+r^2\right)^{5/2}}+\frac{3 a_\Sigma m \left(3 a^2+2 r^2\right)}{2 \left(a^2+r^2\right)^2}-\frac{a_\Sigma  \left(5 a^2+9 r^2\right)}{3 \left(a^2+r^2\right)^{3/2}},\\
    \rho &=& \frac{a^2 }{\left(a^2+r^2\right)^2}+\frac{3 a_\Sigma m \left(a^2+4 r^2\right)}{4 \left(a^2+r^2\right)^2}+\frac{2 a_\Sigma \left(a^2-3 r^2\right)}{3 \left(a^2+r^2\right)^{3/2}}.
\end{eqnarray}
The corrections introduced by this $f(R)$ theory model to the energy conditions are represented by the terms involving $a_\Sigma$. To better understand these modifications, we will analyze the behavior of these combinations graphically.

In Fig \ref{fig:EC_model4}, we graphically analyze how the combinations of the components of the energy-momentum tensor behave as functions of the radial coordinate. The combinations exhibit symmetry under the transformation $r \to -r$, just like the chosen $f(R)$ model itself. For $a_\Sigma = 1$, all conditions, except for $NEC_1$, are satisfied in at least some regions of space-time. For $a_\Sigma = -1$, the $NEC_1$ condition becomes satisfied. However, the other conditions are violated in regions where they were satisfied in the case of GR. However, with the exception of the $SEC_3$ condition, all conditions are satisfied in at least some regions of space-time.

Thus, we conclude that the modifications introduced by the all model of the $f(R)$ theory that we have used can relax the violation of the energy conditions in some regions of space-time. Due to the large amount of information presented in this section, Table \ref{tab:energy_conditions} summarizes the conclusions regarding the energy conditions. The table indicates which inequality is satisfied for each distinct case, considering different values of the constants. The acronyms AS, PS, AV, and PS($r=0$) in the table represent, respectively, Always Satisfied, Partially Satisfied, Always Violated, and Partially Satisfied only in the wormhole throat.

\begin{table}[ht]
\begin{tabular}{c|ccc|ccc|ccc|ccclll}
\cline{1-13}
        & \multicolumn{3}{c|}{\textbf{Case I: $H_R =a_RR^2$}}                                               & \multicolumn{3}{c|}{\textbf{Case II:$H_R=a_1r$}}                                             & \multicolumn{3}{c|}{\textbf{Case III:$H_R=a_2r^2$}}                                              & \multicolumn{3}{c}{\textbf{Case IV:$H_R=a_\Sigma\Sigma$}}                                                           &  &  &  \\ \cline{1-13}
        & \multicolumn{1}{c|}{$a_R=1$} & \multicolumn{1}{c|}{$a_R=0$}   & $a_R=-1$  & \multicolumn{1}{c|}{$a_1=1$} & \multicolumn{1}{c|}{$a_1=0$}   & $a_1=-1$ & \multicolumn{1}{c|}{$a_2=1$}   & \multicolumn{1}{c|}{$a_2=0$}   & $a_2=-1$ & \multicolumn{1}{c|}{$a_\Sigma=1$} & \multicolumn{1}{c|}{$a_\Sigma=0$} & $a_\Sigma=-1$ &  &  &  \\ \cline{1-13}
$NEC_1$ & \multicolumn{1}{c|}{PS}      & \multicolumn{1}{c|}{PS($r=0$)} & PS($r=0$) & \multicolumn{1}{c|}{PS}      & \multicolumn{1}{c|}{PS($r=0$)} & PS       & \multicolumn{1}{c|}{PS($r=0$)} & \multicolumn{1}{c|}{PS($r=0$)} & AS       & \multicolumn{1}{c|}{PS($r=0$)}    & \multicolumn{1}{c|}{PS($r=0$)}    & AS            &  &  &  \\
$NEC_2$ & \multicolumn{1}{c|}{PS}      & \multicolumn{1}{c|}{AS}        & AS        & \multicolumn{1}{c|}{PS}      & \multicolumn{1}{c|}{AS}        & PS       & \multicolumn{1}{c|}{PS}        & \multicolumn{1}{c|}{AS}        & PS       & \multicolumn{1}{c|}{PS}           & \multicolumn{1}{c|}{AS}           & PS            &  &  &  \\
$SEC_3$ & \multicolumn{1}{c|}{PS}      & \multicolumn{1}{c|}{AS}        & AS        & \multicolumn{1}{c|}{PS}      & \multicolumn{1}{c|}{AS}        & PS       & \multicolumn{1}{c|}{AS}        & \multicolumn{1}{c|}{AS}        & PS       & \multicolumn{1}{c|}{AS}           & \multicolumn{1}{c|}{AS}           & AV            &  &  &  \\
$DEC_1$ & \multicolumn{1}{c|}{PS}      & \multicolumn{1}{c|}{PS($r=0$)} & PS($r=0$) & \multicolumn{1}{c|}{AV}      & \multicolumn{1}{c|}{PS($r=0$)} & PS       & \multicolumn{1}{c|}{PS($r=0$)} & \multicolumn{1}{c|}{PS($r=0$)} & AS       & \multicolumn{1}{c|}{PS($r=0$)}    & \multicolumn{1}{c|}{PS($r=0$)}    & PS            &  &  &  \\
$DEC_2$ & \multicolumn{1}{c|}{PS}      & \multicolumn{1}{c|}{PS}        & PS        & \multicolumn{1}{c|}{AV}      & \multicolumn{1}{c|}{PS}        & PS       & \multicolumn{1}{c|}{PS}        & \multicolumn{1}{c|}{PS}        & PS       & \multicolumn{1}{c|}{PS}           & \multicolumn{1}{c|}{PS}           & PS            &  &  &  \\
$WEC_3$ & \multicolumn{1}{c|}{PS}      & \multicolumn{1}{c|}{AS}        & PS        & \multicolumn{1}{c|}{AV}      & \multicolumn{1}{c|}{AS}        & AS       & \multicolumn{1}{c|}{PS}        & \multicolumn{1}{c|}{AS}        & AS       & \multicolumn{1}{c|}{PS}           & \multicolumn{1}{c|}{PS}           & PS            &  &  &  \\ \cline{1-13}
\end{tabular}
\caption{Summary of the energy conditions for the four different cases analyzed. 
The acronyms indicate the status of each condition: AS (Always Satisfied), PS (Partially Satisfied), 
and AV (Always Violated). The notation PS($r=0$) refers to Partial Satisfaction only at the wormhole throat. As in the graphs, for this table we take $a = 2m$, which corresponds to the throat and the horizon being located at $r = 0$.}
\label{tab:energy_conditions}
\end{table}

\section{Conclusions and final remarks}\label{S:conclusions}
In this work, we explore the possibility of generalizing BB solutions, which are well studied in GR, to $f(R)$ gravity by analyzing the source fields that generate these solutions. Our approach involves imposing a well-known space-time, such as the SV metric, within different models of $f(R)$ gravity to determine the necessary source fields for these metrics to be solutions of the field equations.  

As our main result, we show that, given the stress-energy tensor $T^{(GR)}_{\mu\nu}$ in GR, we can always determine the corresponding stress-energy tensor $T^{(H)}_{\mu\nu}$ in $f(R)$ gravity when $f(R) = R + H(R)$. However, this is only possible if we consider a magnetic source within nonlinear electrodynamics. In the case of an electric source, the stress-energy tensor cannot be split, and the approach does not apply. We obtain general analytical expressions for the functions $L^{(H)}$, $L_F^{(H)}$, $V^{(H)}$, and $h^{(H)}$, which depend on the forms of the functions $H_R$, $H(R)$, $A(r)$, and $\Sigma(r)$. Thus, it becomes evident that the chosen form of the gravity theory and the metric will influence the form of the functions related to the source fields.

As a first case, we considered the Starobinsky model, $f(R) = R + a_R R^2$, i.e, $H(R) = a_RR^2$, which is widely studied in the literature. We showed that if the constant $a_R$ is positive, $h(\phi)$ can take positive values for $r \to 0$, provided that $a > 9m/2$. This indicates that, unlike GR, the scalar field does not always have to be a phantom. The second approach chosen was $H_R =  a_1 r$, inspired by the form obtained for regular BHs in other works. In this second model, the function $h(\phi)$ is simpler than in the first case, but if the constant $a_1$ is positive, the scalar field will always be a phantom, offering no advantages compared to GR. For the scalar field to be canonical in some regions, we must assume that the constant $a_1$ can take negative values, however, there will still be regions where the scalar field remains phantom. The functions related to the source fields and the $f(R)$ theory are not symmetric under the transformation $r \to -r$, implying that the function $f(R)$ is multivalued.
We chose the model $H_R = a_2 r^2$ for the third case. The advantage of this model is that the source fields and the function $f(R)$ are symmetric under the transformation $r \to -r$, so unlike the previous case, the function $f(R)$ will not be multivalued. If the constant $a_2$ is positive, then the scalar field will always be a phantom. However, for negative $a_2$, if $a^2 > -1/a_2$, our scalar field will always be canonical. This is the first case in which we find a BB solution where the scalar field can always be canonical, depending on the choice of parameters. For the last model, we considered $H_R =  a_\Sigma \Sigma$. This proposal is a modification of the second model by replacing $r \to \sqrt{r^2 + a^2}$. Interestingly, this model yields more simplified solutions for the source fields. As in the third model, we obtain symmetry under the transformation $r \to -r$ in all functions derived for the source field and the function $f(R)$. This is the second case where the scalar field can always be canonical, provided that the constant $a_\Sigma$ takes negative values and $a$ is sufficiently large. Thus, we have constructed at least two models for generating the BB solution through a canonical scalar field.

We also analyzed the energy conditions, and one general feature across all models is that, depending on the chosen parameters, each of the inequalities will be satisfied in specific regions of space-time. In cases I and III, regions exist where all energy conditions are satisfied. However, this is not true in cases II and IV, as at least one inequality will always be violated, violating some energy conditions. Overall, including the $f(R)$ theory relaxes the necessity of energy condition violations, allowing for a much broader range of scenarios where these conditions can be satisfied compared to GR.

It is important to highlight that, although our geometric ansatz is fixed , that is, the SV metric, the introduction of different $f(R)$ theories and, consequently, of different NED sources that support it, have interesting observable implications. The presence of a NED source implies that photon trajectories no longer follow the standard spacetime metric \cite{Novello2000,Novello2001,Silva:2024fpn}, but rather those of an $
g^{\mu\nu}_{\rm eff} = L_Fg^{\mu\nu} - L_{FF}F^{\mu\lambda}F_\lambda{}^\nu $, which, in our case, where $L(F) = L^{(GR)}(F) + L^{(H)}(F)$, can be rewritten as $g^{\mu\nu}_{\rm eff} = \,\,^{(GR)}g^{\mu\nu}_{\rm eff} + ^{(H)}g^{\mu\nu}_{\rm eff}$.
In the spherically symmetric BB geometry used here, this means that the photon sphere and circular photon orbits are shifted with respect to the “naïve” geodesic structure of the background metric.  
Since the shadow of the compact object (and related photon ring/lensing observables) is determined essentially by the unstable photon orbits, the modification induced by NED can lead to measurable deviations in the shadow size, shape and brightness profile when compared to GR case analyzed in \cite{Silva:2024fpn}.
Therefore, from an observational standpoint, the corrections to the photon trajectories introduced by NED in our model admit a direct link to astrophysical data by computing the modified photon sphere radius and tracing backward the corresponding shadow contour, one can in principle compare with e.g. the EHT measurements to place constraints on the strength or form of the NED coupling.  
In summary, the incorporation of photon trajectories in the effective metric opens a pathway to connect the theoretical features of the $f(R) +$ NED BB model with optical/astrophysical observables, thus enhancing the testability of the model in the strong-gravity regime.

 Moreover, using the $f(R)$ theory requires the introduction of additional nonlinearities in both the scalar and electromagnetic sectors, which are essential for studying stability through scalar and electromagnetic perturbations. Moreover, modifications in nonlinear electrodynamics influence photon trajectories in this space-time and affect the thermodynamics of BHs. Beyond this, the approach developed here can be applied to other BB solutions, such as the Reissner-Nordström, BTZ, black strings, and many others found in the literature. All these aspects will be explored in future work, and thus, this study opens new possibilities for further research. 
\section*{Acknowledgments}
\hspace{0.5cm} The authors would like to thank Conselho Nacional de Desenvolvimento Cient\'{i}fico e Tecnol\'ogico (CNPq) and Funda\c c\~ao Cearense de Apoio ao Desenvolvimento Cient\'ifico e Tecnol\'ogico (FUNCAP) for the financial support.

\bibliographystyle{apsrev4-1}
\bibliography{ref.bib}

@article{Simpson:2018tsi,
    author = "Simpson, Alex and Visser, Matt",
    title = "{Black-bounce to traversable wormhole}",
    eprint = "1812.07114",
    archivePrefix = "arXiv",
    primaryClass = "gr-qc",
    doi = "10.1088/1475-7516/2019/02/042",
    journal = "JCAP",
    volume = "02",
    pages = "042",
    year = "2019"
}

@article{LIGOScientific:2016aoc,
    author = "Abbott, B. P. and others",
    collaboration = "LIGO Scientific, Virgo",
    title = "{Observation of Gravitational Waves from a Binary Black Hole Merger}",
    eprint = "1602.03837",
    archivePrefix = "arXiv",
    primaryClass = "gr-qc",
    reportNumber = "LIGO-P150914",
    doi = "10.1103/PhysRevLett.116.061102",
    journal = "Phys. Rev. Lett.",
    volume = "116",
    number = "6",
    pages = "061102",
    year = "2016"
}

@article{Rodrigues:2022mdm,
    author = "Rodrigues, Manuel E. and Silva, Marcos V. de S.",
    title = "{Black-bounces with multiple throats and anti-throats}",
    eprint = "2204.11851",
    archivePrefix = "arXiv",
    primaryClass = "gr-qc",
    doi = "10.1088/1361-6382/ad0195",
    journal = "Class. Quant. Grav.",
    volume = "40",
    number = "22",
    pages = "225011",
    year = "2023"
}

@article{Rodrigues:2022rfj,
    author = "Rodrigues, Manuel E. and Silva, Marcos V. de S.",
    title = "{Embedding regular black holes and black bounces in a cloud of strings}",
    eprint = "2210.05383",
    archivePrefix = "arXiv",
    primaryClass = "gr-qc",
    doi = "10.1103/PhysRevD.106.084016",
    journal = "Phys. Rev. D",
    volume = "106",
    number = "8",
    pages = "084016",
    year = "2022"
}

@article{EventHorizonTelescope:2019dse,
    author = "Akiyama, Kazunori and others",
    collaboration = "Event Horizon Telescope",
    title = "{First M87 Event Horizon Telescope Results. I. The Shadow of the Supermassive Black Hole}",
    eprint = "1906.11238",
    archivePrefix = "arXiv",
    primaryClass = "astro-ph.GA",
    doi = "10.3847/2041-8213/ab0ec7",
    journal = "Astrophys. J. Lett.",
    volume = "875",
    pages = "L1",
    year = "2019"
}

@book{Bronnikov:2012wsj,
    author = "Bronnikov, Kirill A. and Rubin, Sergey G.",
    title = "{Black Holes, Cosmology and Extra Dimensions}",
    doi = "10.1142/12186",
    isbn = "978-981-4374-20-0, 978-981-4440-02-8",
    publisher = "WSP",
    year = "2012"
}

@article{Lima:2021las,
    author = "Lima, Junior., Haroldo C. D. and Crispino, Lu\'\i{}s C. B. and Cunha, Pedro V. P. and Herdeiro, Carlos A. R.",
    title = "{Can different black holes cast the same shadow?}",
    eprint = "2102.07034",
    archivePrefix = "arXiv",
    primaryClass = "gr-qc",
    doi = "10.1103/PhysRevD.103.084040",
    journal = "Phys. Rev. D",
    volume = "103",
    number = "8",
    pages = "084040",
    year = "2021"
}

@article{Lima:2022pvc,
    author = "Lima, Arthur Menezes and de Alencar Filho, Geov\'a Maciel and Furtado Neto, Job Saraiva",
    title = "{Black String Bounce to Traversable Wormhole}",
    eprint = "2211.12349",
    archivePrefix = "arXiv",
    primaryClass = "gr-qc",
    doi = "10.3390/sym15010150",
    journal = "Symmetry",
    volume = "15",
    number = "1",
    pages = "150",
    year = "2023"
}

@article{Franzin:2023slm,
    author = "Franzin, Edgardo and Liberati, Stefano and Vellucci, Vania",
    title = "{From regular black holes to horizonless objects: quasi-normal modes, instabilities and spectroscopy}",
    eprint = "2310.11990",
    archivePrefix = "arXiv",
    primaryClass = "gr-qc",
    doi = "10.1088/1475-7516/2024/01/020",
    journal = "JCAP",
    volume = "01",
    pages = "020",
    year = "2024"
}

@article{Silva:2024fpn,
    author = "Silva, Marcos V. de S. and Rodrigues, Manuel E.",
    title = "{Orbits Around a Black Bounce Spacetime}",
    eprint = "2404.15792",
    archivePrefix = "arXiv",
    primaryClass = "gr-qc",
    doi = "10.1007/s10773-024-05644-5",
    journal = "Int. J. Theor. Phys.",
    volume = "63",
    number = "4",
    pages = "101",
    year = "2024"
}

@article{Bronnikov:2021uta,
    author = "Bronnikov, Kirill A. and Walia, Rahul Kumar",
    title = "{Field sources for Simpson-Visser spacetimes}",
    eprint = "2112.13198",
    archivePrefix = "arXiv",
    primaryClass = "gr-qc",
    doi = "10.1103/PhysRevD.105.044039",
    journal = "Phys. Rev. D",
    volume = "105",
    number = "4",
    pages = "044039",
    year = "2022"
}

@article{dePaula:2024yzy,
    author = "de Paula, Marco A. A. and Lima, Junior., Haroldo C. D. and Cunha, Pedro V. P. and Herdeiro, Carlos A. R. and Crispino, Lu\'\i{}s C. B.",
    title = "{Good tachyons, bad bradyons: Role reversal in Einstein-nonlinear-electrodynamics models}",
    eprint = "2412.18659",
    archivePrefix = "arXiv",
    primaryClass = "gr-qc",
    doi = "10.1016/j.physletb.2025.139513",
    journal = "Phys. Lett. B",
    volume = "866",
    pages = "139513",
    year = "2025"
}

@article{Rodrigues:2022qdp,
    author = "Rodrigues, Manuel E. and de S. Silva, Marcos V. and Vieira, Henrique A.",
    title = "{Bardeen-Kiselev black hole with a cosmological constant}",
    eprint = "2203.04965",
    archivePrefix = "arXiv",
    primaryClass = "gr-qc",
    doi = "10.1103/PhysRevD.105.084043",
    journal = "Phys. Rev. D",
    volume = "105",
    number = "8",
    pages = "084043",
    year = "2022"
}

@article{Canate:2022gpy,
    author = "Ca\~nate, Pedro",
    title = "{Black bounces as magnetically charged phantom regular black holes in Einstein-nonlinear electrodynamics gravity coupled to a self-interacting scalar field}",
    eprint = "2202.02303",
    archivePrefix = "arXiv",
    primaryClass = "gr-qc",
    doi = "10.1103/PhysRevD.106.024031",
    journal = "Phys. Rev. D",
    volume = "106",
    number = "2",
    pages = "024031",
    year = "2022"
}

@article{Rodrigues:2023vtm,
    author = "Rodrigues, Manuel E. and Silva, Marcos V. de S.",
    title = "{Source of black bounces in general relativity}",
    eprint = "2302.10772",
    archivePrefix = "arXiv",
    primaryClass = "gr-qc",
    doi = "10.1103/PhysRevD.107.044064",
    journal = "Phys. Rev. D",
    volume = "107",
    number = "4",
    pages = "044064",
    year = "2023"
}

@article{Pereira:2024rtv,
    author = "Pereira, Carlos F. S. and C. Rodrigues, Denis and Silva, Marcos V. de S. and Fabris, J\'ulio C. and Rodrigues, Manuel E. and Belich, H.",
    title = "{Magnetically charged black-bounce solution via nonlinear electrodynamics in a k-essence theory}",
    eprint = "2409.09182",
    archivePrefix = "arXiv",
    primaryClass = "gr-qc",
    doi = "10.1103/PhysRevD.111.084025",
    journal = "Phys. Rev. D",
    volume = "111",
    number = "8",
    pages = "084025",
    year = "2025"
}

@article{Bronnikov:2022bud,
    author = "Bronnikov, K. A.",
    title = "{Black bounces, wormholes, and partly phantom scalar fields}",
    eprint = "2206.09227",
    archivePrefix = "arXiv",
    primaryClass = "gr-qc",
    doi = "10.1103/PhysRevD.106.064029",
    journal = "Phys. Rev. D",
    volume = "106",
    number = "6",
    pages = "064029",
    year = "2022"
}

@article{Alencar:2024nxi,
    author = "Alencar, G. and Nilton, M. and Rodrigues, Manuel E. and de S. Silva, Marcos V.",
    title = "{Field sources for f(R,R{\ensuremath{\mu}}{\ensuremath{\nu}}) black-bounce solutions: The case of K-gravity}",
    eprint = "2409.12101",
    archivePrefix = "arXiv",
    primaryClass = "gr-qc",
    doi = "10.1016/j.dark.2025.102060",
    journal = "Phys. Dark Univ.",
    volume = "49",
    pages = "102060",
    year = "2025"
}

@article{Starobinsky:1980te,
    author = "Starobinsky, Alexei A.",
    editor = "Khalatnikov, I. M. and Mineev, V. P.",
    title = "{A New Type of Isotropic Cosmological Models Without Singularity}",
    doi = "10.1016/0370-2693(80)90670-X",
    journal = "Phys. Lett. B",
    volume = "91",
    pages = "99--102",
    year = "1980"
}

@article{DeFelice:2010aj,
    author = "De Felice, Antonio and Tsujikawa, Shinji",
    title = "{f(R) theories}",
    eprint = "1002.4928",
    archivePrefix = "arXiv",
    primaryClass = "gr-qc",
    doi = "10.12942/lrr-2010-3",
    journal = "Living Rev. Rel.",
    volume = "13",
    pages = "3",
    year = "2010"
}

@article{Rodrigues:2015ayd,
    author = "Rodrigues, Manuel E. and Junior, Ednaldo L. B. and Marques, Glauber T. and Zanchin, Vilson T.",
    title = "{Regular black holes in $f(R)$ gravity coupled to nonlinear electrodynamics}",
    eprint = "1511.00569",
    archivePrefix = "arXiv",
    primaryClass = "gr-qc",
    doi = "10.1103/PhysRevD.94.024062",
    journal = "Phys. Rev. D",
    volume = "94",
    number = "2",
    pages = "024062",
    year = "2016",
    note = "[Addendum: Phys.Rev.D 94, 049904 (2016)]"
}

@book{Visser:1995cc,
    author = "Visser, Matt",
    title = "{Lorentzian wormholes: From Einstein to Hawking}",
    isbn = "978-1-56396-653-8",
    year = "1995"
}

@article{Lima:2023arg,
    author = "Lima, A. and Alencar, G. and Costa Filho, R. N. and Landim, R. R.",
    title = "{Charged black string bounce and its field source}",
    eprint = "2306.03029",
    archivePrefix = "arXiv",
    primaryClass = "gr-qc",
    doi = "10.1007/s10714-023-03156-x",
    journal = "Gen. Rel. Grav.",
    volume = "55",
    number = "10",
    pages = "108",
    year = "2023"
}

@article{Lima:2023jtl,
    author = "Lima, A. and Alencar, G. and S\'aez-Chillon G\'omez, Diego",
    title = "{Regularizing rotating black strings with a new black-bounce solution}",
    eprint = "2307.07404",
    archivePrefix = "arXiv",
    primaryClass = "gr-qc",
    doi = "10.1103/PhysRevD.109.064038",
    journal = "Phys. Rev. D",
    volume = "109",
    number = "6",
    pages = "064038",
    year = "2024"
}

@article{Crispim:2024yjz,
    author = "Crispim, Tiago M. and Estrada, Milko and Muniz, C. R. and Alencar, G.",
    title = "{Braneworld black bounce to transversable wormhole}",
    eprint = "2405.08048",
    archivePrefix = "arXiv",
    primaryClass = "hep-th",
    doi = "10.1088/1475-7516/2024/10/063",
    journal = "JCAP",
    volume = "10",
    pages = "063",
    year = "2024"
}

@article{Crispim:2024nou,
    author = "Crispim, T. M. and Alencar, G. and Estrada, Milko",
    title = "{Braneworld Black Bounce to Transversable Wormhole Analytically Connected to an asymptotically $AdS_5$ Boundary}",
    eprint = "2407.03528",
    archivePrefix = "arXiv",
    primaryClass = "gr-qc",
    month = "7",
    year = "2024"
}

@article{Islam:2021ful,
    author = "Islam, Shafqat Ul and Kumar, Jitendra and Ghosh, Sushant G.",
    title = "{Strong gravitational lensing by
rotating Simpson-Visser black holes}",
    eprint = "2104.00696",
    archivePrefix = "arXiv",
    primaryClass = "gr-qc",
    doi = "10.1088/1475-7516/2021/10/013",
    journal = "JCAP",
    volume = "10",
    pages = "013",
    year = "2021"
}

@article{Nascimento:2020ime,
    author = "Nascimento, J. R. and Petrov, A. Yu. and Porfirio, P. J. and Soares, A. R.",
    title = "{Gravitational lensing in black-bounce spacetimes}",
    eprint = "2005.13096",
    archivePrefix = "arXiv",
    primaryClass = "gr-qc",
    doi = "10.1103/PhysRevD.102.044021",
    journal = "Phys. Rev. D",
    volume = "102",
    number = "4",
    pages = "044021",
    year = "2020"
}

@article{Ghosh:2022mka,
    author = "Ghosh, Saptaswa and Bhattacharyya, Arpan",
    title = "{Analytical study of gravitational lensing in Kerr-Newman black-bounce spacetime}",
    eprint = "2206.09954",
    archivePrefix = "arXiv",
    primaryClass = "gr-qc",
    doi = "10.1088/1475-7516/2022/11/006",
    journal = "JCAP",
    volume = "11",
    pages = "006",
    year = "2022"
}

@article{Guerrero:2021ues,
    author = "Guerrero, Merce and Olmo, Gonzalo J. and Rubiera-Garcia, Diego and G\'omez, Diego S\'aez-Chill\'on",
    title = "{Shadows and optical appearance of black bounces illuminated by a thin accretion disk}",
    eprint = "2105.15073",
    archivePrefix = "arXiv",
    primaryClass = "gr-qc",
    doi = "10.1088/1475-7516/2021/08/036",
    journal = "JCAP",
    volume = "08",
    pages = "036",
    year = "2021"
}

@article{Guo:2021wid,
    author = "Guo, Yang and Miao, Yan-Gang",
    title = "{Charged black-bounce spacetimes: Photon rings, shadows and observational appearances}",
    eprint = "2112.01747",
    archivePrefix = "arXiv",
    primaryClass = "gr-qc",
    doi = "10.1016/j.nuclphysb.2022.115938",
    journal = "Nucl. Phys. B",
    volume = "983",
    pages = "115938",
    year = "2022"
}

@article{Olmo:2023lil,
    author = "Olmo, Gonzalo J. and Rosa, Joao Luis and Rubiera-Garcia, Diego and Saez-Chillon Gomez, Diego",
    title = "{Shadows and photon rings of regular black holes and geonic horizonless compact objects}",
    eprint = "2302.12064",
    archivePrefix = "arXiv",
    primaryClass = "gr-qc",
    reportNumber = "IPARCOS-UCM-23-111",
    doi = "10.1088/1361-6382/aceacd",
    journal = "Class. Quant. Grav.",
    volume = "40",
    number = "17",
    pages = "174002",
    year = "2023"
}

@article{Bronnikov:2023aya,
    author = "Bronnikov, Kirill A. and Rodrigues, Manuel E. and de S. Silva, Marcos V.",
    title = "{Cylindrical black bounces and their field sources}",
    eprint = "2305.19296",
    archivePrefix = "arXiv",
    primaryClass = "gr-qc",
    doi = "10.1103/PhysRevD.108.024065",
    journal = "Phys. Rev. D",
    volume = "108",
    number = "2",
    pages = "024065",
    year = "2023"
}

@article{Yang:2021cvh,
    author = "Yang, Yi and Liu, Dong and Xu, Zhaoyi and Xing, Yujia and Wu, Shurui and Long, Zheng-Wen",
    title = "{Echoes of novel black-bounce spacetimes}",
    eprint = "2107.06554",
    archivePrefix = "arXiv",
    primaryClass = "gr-qc",
    doi = "10.1103/PhysRevD.104.104021",
    journal = "Phys. Rev. D",
    volume = "104",
    number = "10",
    pages = "104021",
    year = "2021"
}

@article{Franzin:2022iai,
    author = "Franzin, Edgardo and Liberati, Stefano and Mazza, Jacopo and Dey, Ramit and Chakraborty, Sumanta",
    title = "{Scalar perturbations around rotating regular black holes and wormholes: Quasinormal modes, ergoregion instability, and superradiance}",
    eprint = "2201.01650",
    archivePrefix = "arXiv",
    primaryClass = "gr-qc",
    doi = "10.1103/PhysRevD.105.124051",
    journal = "Phys. Rev. D",
    volume = "105",
    number = "12",
    pages = "124051",
    year = "2022"
}

@inproceedings{bardeen1968,
  author = {Bardeen, J. M.},
  title = {Non-singular general-relativistic gravitational collapse},
  booktitle = {Proceedings of the International Conference on Gravitation and Collapse},
  year = {1968},
  address = {Tbilisi, USSR}
}

@article{hayward2006,
  author = {Hayward, S. A.},
  title = {Formation and evaporation of non-singular black holes},
  journal = {Physical Review Letters},
  volume = {96},
  pages = {031103},
  year = {2006},
  doi = {10.1103/PhysRevLett.96.031103}
}

@article{dymnikova1992,
  author = {Dymnikova, I. G.},
  title = {Vacuum nonsingular black hole},
  journal = {General Relativity and Gravitation},
  volume = {24},
  pages = {235-242},
  year = {1992},
  doi = {10.1007/BF00760226}
}

@article{ellis1973,
  author = {Ellis, H. G.},
  title = {Ether flow through a drainhole: A particle model in general relativity},
  journal = {Journal of Mathematical Physics},
  volume = {14},
  pages = {104-118},
  year = {1973},
  doi = {10.1063/1.1666161}
}

@article{bronnikov1973,
  author = {Bronnikov, K. A.},
  title = {Scalar-tensor theory and scalar charge},
  journal = {Acta Physica Polonica B},
  volume = {4},
  pages = {251-266},
  year = {1973},
  doi = {10.1007/BF00760226}
}

@article{morris1988,
  author = {Morris, M. S. and Thorne, K. S.},
  title = {Wormholes, time machines, and the weak energy condition},
  journal = {Physical Review Letters},
  volume = {61},
  pages = {1446-1449},
  year = {1988},
  doi = {10.1103/PhysRevLett.61.1446}
}

@article{Rodrigues:2018bdc,
    author = "Rodrigues, Manuel E. and de Sousa Silva, Marcos V.",
    title = "{Bardeen Regular Black Hole With an Electric Source}",
    eprint = "1802.05095",
    archivePrefix = "arXiv",
    primaryClass = "gr-qc",
    doi = "10.1088/1475-7516/2018/06/025",
    journal = "JCAP",
    volume = "06",
    pages = "025",
    year = "2018"
}

@article{Ayon-Beato:2000mjt,
    author = "Ayon-Beato, Eloy and Garcia, Alberto",
    title = "{The Bardeen model as a nonlinear magnetic monopole}",
    eprint = "gr-qc/0009077",
    archivePrefix = "arXiv",
    doi = "10.1016/S0370-2693(00)01125-4",
    journal = "Phys. Lett. B",
    volume = "493",
    pages = "149--152",
    year = "2000"
}

@article{Bronnikov:2018vbs,
    author = "Bronnikov, Kirill A.",
    editor = "Sedrakian, Armen",
    title = "{Scalar fields as sources for wormholes and regular black holes}",
    eprint = "1802.00098",
    archivePrefix = "arXiv",
    primaryClass = "gr-qc",
    doi = "10.3390/particles1010005",
    journal = "Particles",
    volume = "1",
    number = "1",
    pages = "56--81",
    year = "2018"
}

@article{Crispim:2024dgd,
    author = "Crispim, T. M. and Alencar, G. and Muniz, C. R.",
    title = "{Field Sources for Generalized Ellis-Bronnikov Wormhole}",
    eprint = "2410.11147",
    archivePrefix = "arXiv",
    primaryClass = "gr-qc",
    month = "10",
    year = "2024"
}

@article{deSSilva:2024gdc,
    author = "de S. Silva, Marcos V. and Alencar, G. and Costa Filho, R. N. and Neves, R. M. P. and Muniz, Celio R.",
    title = "{Traversable wormholes sourced by dark matter in loop quantum cosmology}",
    eprint = "2411.12063",
    archivePrefix = "arXiv",
    primaryClass = "gr-qc",
    doi = "10.1140/epjp/s13360-025-06214-2",
    journal = "Eur. Phys. J. Plus",
    volume = "140",
    number = "4",
    pages = "289",
    year = "2025"
}

@article{Crispim:2024lzf,
    author = "Crispim, T. M. and Silva, Marcos V. de S. and Alencar, G. and Muniz, Celio R. and S\'aez-Chill\'on G\'omez, Diego",
    title = "{Field Sources for Wormholes With Multiple Throats/Anti-throats}",
    eprint = "2412.05236",
    archivePrefix = "arXiv",
    primaryClass = "gr-qc",
    doi = "10.1088/1361-6382/adc654",
    journal = "Class. Quant. Grav.",
    volume = "42",
    pages = "085005",
    year = "2025"
}

@article{Alencar:2024yvh,
    author = "Alencar, G. and Bronnikov, Kirill A. and Rodrigues, Manuel E. and S\'aez-Chill\'on G\'omez, Diego and de S. Silva, Marcos V.",
    title = "{On black bounce space-times in non-linear electrodynamics}",
    eprint = "2403.12897",
    archivePrefix = "arXiv",
    primaryClass = "gr-qc",
    doi = "10.1140/epjc/s10052-024-13119-4",
    journal = "Eur. Phys. J. C",
    volume = "84",
    number = "7",
    pages = "745",
    year = "2024"
}

@article{Alencar:2025jvl,
    author = "Alencar, G. and Duran-Cabac\'es, Albert and Rubiera-Garcia, Diego and S\'aez-Chill\'on G\'omez, Diego",
    title = "{General spherically symmetric black bounces within nonlinear electrodynamics}",
    eprint = "2501.03909",
    archivePrefix = "arXiv",
    primaryClass = "gr-qc",
    doi = "10.1103/PhysRevD.111.104020",
    journal = "Phys. Rev. D",
    volume = "111",
    number = "10",
    pages = "104020",
    year = "2025"
}

@article{Pereira:2024gsl,
    author = "Pereira, Carlos F. S. and C. Rodrigues, Denis and Martins, \'Ebano L. and Fabris, J\'ulio C. and Rodrigues, Manuel E.",
    title = "{New sources of ghost fields in k-essence theories for black-bounce solutions}",
    eprint = "2405.07455",
    archivePrefix = "arXiv",
    primaryClass = "gr-qc",
    doi = "10.1088/1361-6382/ad98e0",
    journal = "Class. Quant. Grav.",
    volume = "42",
    number = "1",
    pages = "015001",
    year = "2025"
}

@article{Capozziello:2007id,
    author = "Capozziello, S. and Stabile, A. and Troisi, A.",
    title = "{Spherical symmetry in f(R)-gravity}",
    eprint = "0709.0891",
    archivePrefix = "arXiv",
    primaryClass = "gr-qc",
    doi = "10.1088/0264-9381/25/8/085004",
    journal = "Class. Quant. Grav.",
    volume = "25",
    pages = "085004",
    year = "2008"
}

@article{Sebastiani:2010kv,
    author = "Sebastiani, Lorenzo and Zerbini, Sergio",
    title = "{Static Spherically Symmetric Solutions in F(R) Gravity}",
    eprint = "1012.5230",
    archivePrefix = "arXiv",
    primaryClass = "gr-qc",
    doi = "10.1140/epjc/s10052-011-1591-8",
    journal = "Eur. Phys. J. C",
    volume = "71",
    pages = "1591",
    year = "2011"
}

@article{Capozziello:2007wc,
    author = "Capozziello, S. and Stabile, A. and Troisi, A.",
    title = "{Spherically symmetric solutions in f(R)-gravity via Noether Symmetry Approach}",
    eprint = "gr-qc/0703067",
    archivePrefix = "arXiv",
    doi = "10.1088/0264-9381/24/8/013",
    journal = "Class. Quant. Grav.",
    volume = "24",
    pages = "2153--2166",
    year = "2007"
}

@article{Kainulainen:2007bt,
    author = "Kainulainen, Kimmo and Piilonen, Johanna and Reijonen, Vappu and Sunhede, Daniel",
    title = "{Spherically symmetric spacetimes in f(R) gravity theories}",
    eprint = "0704.2729",
    archivePrefix = "arXiv",
    primaryClass = "gr-qc",
    doi = "10.1103/PhysRevD.76.024020",
    journal = "Phys. Rev. D",
    volume = "76",
    pages = "024020",
    year = "2007"
}

@article{Nashed:2019tuk,
    author = "Nashed, Gamal G. L. and Capozziello, Salvatore",
    title = "{Charged spherically symmetric black holes in $f(R)$ gravity and their stability analysis}",
    eprint = "1902.06783",
    archivePrefix = "arXiv",
    primaryClass = "gr-qc",
    doi = "10.1103/PhysRevD.99.104018",
    journal = "Phys. Rev. D",
    volume = "99",
    number = "10",
    pages = "104018",
    year = "2019"
}

@article{Multamaki:2006zb,
    author = "Multamaki, Tuomas and Vilja, Iiro",
    title = "{Spherically symmetric solutions of modified field equations in f(R) theories of gravity}",
    eprint = "astro-ph/0606373",
    archivePrefix = "arXiv",
    doi = "10.1103/PhysRevD.74.064022",
    journal = "Phys. Rev. D",
    volume = "74",
    pages = "064022",
    year = "2006"
}

@article{DeBenedictis:2012qz,
    author = "DeBenedictis, Andrew and Horvat, Dubravko",
    title = "{On Wormhole Throats in $f(R)$ Gravity Theory}",
    eprint = "1111.3704",
    archivePrefix = "arXiv",
    primaryClass = "gr-qc",
    doi = "10.1007/s10714-012-1412-x",
    journal = "Gen. Rel. Grav.",
    volume = "44",
    pages = "2711--2744",
    year = "2012"
}

@article{Samanta:2019tjb,
    author = "Samanta, Gauranga C. and Godani, Nisha",
    title = "{Validation of energy conditions in wormhole geometry within viable $f(R)$ gravity}",
    eprint = "1908.04406",
    archivePrefix = "arXiv",
    primaryClass = "gr-qc",
    doi = "10.1140/epjc/s10052-019-7116-6",
    journal = "Eur. Phys. J. C",
    volume = "79",
    number = "7",
    pages = "623",
    year = "2019"
}

@article{Bhattacharya:2015oma,
    author = "Bhattacharya, Subhra and Chakraborty, S.",
    title = "{$f(R)$ gravity solutions for evolving wormholes}",
    eprint = "1506.03968",
    archivePrefix = "arXiv",
    primaryClass = "gr-qc",
    doi = "10.1140/epjc/s10052-017-5131-z",
    journal = "Eur. Phys. J. C",
    volume = "77",
    number = "8",
    pages = "558",
    year = "2017"
}

@article{Bambi:2015zch,
    author = "Bambi, Cosimo and Cardenas-Avendano, Alejandro and Olmo, Gonzalo J. and Rubiera-Garcia, D.",
    title = "{Wormholes and nonsingular spacetimes in Palatini $f(R)$ gravity}",
    eprint = "1511.03755",
    archivePrefix = "arXiv",
    primaryClass = "gr-qc",
    reportNumber = "IFIC-16-58",
    doi = "10.1103/PhysRevD.93.064016",
    journal = "Phys. Rev. D",
    volume = "93",
    number = "6",
    pages = "064016",
    year = "2016"
}

@article{Rodrigues:2016fym,
    author = "Rodrigues, Manuel E. and Fabris, Julio C. and Junior, Ednaldo L. B. and Marques, Glauber T.",
    title = "{Generalisation for regular black holes on general relativity to $f(R)$ gravity}",
    eprint = "1601.00471",
    archivePrefix = "arXiv",
    primaryClass = "gr-qc",
    doi = "10.1140/epjc/s10052-016-4085-x",
    journal = "Eur. Phys. J. C",
    volume = "76",
    number = "5",
    pages = "250",
    year = "2016"
}

@article{Santos:2023vox,
    author = "Santos, A. C. L. and Maluf, R. V. and Muniz, C. R.",
    title = "{Regular Black Holes in $D=2+1$ with $f(R)$ Gravity}",
    eprint = "2312.05326",
    archivePrefix = "arXiv",
    primaryClass = "gr-qc",
    month = "12",
    year = "2023"
}

@article{Rodrigues:2019xrc,
    author = "Rodrigues, Manuel E. and de Sousa Silva, Marcos V.",
    title = "{Regular multihorizon black holes in $f(G)$ gravity with nonlinear electrodynamics}",
    eprint = "1906.06168",
    archivePrefix = "arXiv",
    primaryClass = "gr-qc",
    doi = "10.1103/PhysRevD.99.124010",
    journal = "Phys. Rev. D",
    volume = "99",
    number = "12",
    pages = "124010",
    year = "2019"
}

@article{Lobo:2020ffi,
    author = "Lobo, Francisco S. N. and Rodrigues, Manuel E. and de Sousa Silva, Marcos V. and Simpson, Alex and Visser, Matt",
    title = "{Novel black-bounce spacetimes: wormholes, regularity, energy conditions, and causal structure}",
    eprint = "2009.12057",
    archivePrefix = "arXiv",
    primaryClass = "gr-qc",
    doi = "10.1103/PhysRevD.103.084052",
    journal = "Phys. Rev. D",
    volume = "103",
    number = "8",
    pages = "084052",
    year = "2021"
}

@article{Rois:2024iiu,
    author = "R{\'o}is, Gabriel I. and Junior, Jos{\'e} Tarciso S. S. and Lobo, Francisco S. N. and Rodrigues, Manuel E. and Harko, Tiberiu",
    title = "{Charged black hole solutions in f(R,T) gravity coupled to nonlinear electrodynamics}",
    eprint = "2412.00582",
    archivePrefix = "arXiv",
    primaryClass = "gr-qc",
    doi = "10.1103/srr5-t81h",
    journal = "Phys. Rev. D",
    volume = "111",
    number = "12",
    pages = "124044",
    year = "2025"
}

@article{Tangphati:2023xnw,
    author = "Tangphati, Takol and Youk, Menglong and Ponglertsakul, Supakchai",
    title = "{Magnetically charged regular black holes in f(R,T) gravity coupled to nonlinear electrodynamics}",
    eprint = "2312.16614",
    archivePrefix = "arXiv",
    primaryClass = "gr-qc",
    doi = "10.1016/j.jheap.2024.06.009",
    journal = "JHEAp",
    volume = "43",
    pages = "66--78",
    year = "2024"
}

@article{Rodrigues:2023fps,
    author = "Rodrigues, Manuel E. and Vieira, Henrique A.",
    title = "{A regular metric does not ensure the regularity of spacetime}",
    eprint = "2311.02138",
    archivePrefix = "arXiv",
    primaryClass = "gr-qc",
    doi = "10.1140/epjp/s13360-023-04624-8",
    journal = "Eur. Phys. J. Plus",
    volume = "138",
    number = "11",
    pages = "974",
    year = "2023"
}

@article{Pal:2023rvv,
    author = "Pal, Kunal and Pal, Kuntal and Sarkar, Tapobrata",
    title = "{Geodesically completing regular black holes by the Simpson\textendash{}Visser method}",
    eprint = "2307.09382",
    archivePrefix = "arXiv",
    primaryClass = "gr-qc",
    doi = "10.1007/s10714-023-03168-7",
    journal = "Gen. Rel. Grav.",
    volume = "55",
    number = "10",
    pages = "121",
    year = "2023"
}

@article{Rodrigues:2025plw,
    author = "Rodrigues, Manuel E. and Silva, Marcos V. de S.",
    title = "{Spherically symmetric and static black bounces with multiple horizons, throats, and anti-throats in four dimensions}",
    eprint = "2502.00502",
    archivePrefix = "arXiv",
    primaryClass = "gr-qc",
    doi = "10.1088/1361-6382/adaf01",
    journal = "Class. Quant. Grav.",
    volume = "42",
    number = "5",
    pages = "055005",
    year = "2025"
}

@article{Junior:2025sjr,
    author = "Junior, Ednaldo L. B. and Junior, Jos{\'e} Tarciso S. S. and Lobo, Francisco S. N. and Rodrigues, Manuel E. and da Silva, Lu{\'\i}s F. Dias and Vieira, Henrique A.",
    title = "{Dyonic regular black bounce solutions in general relativity}",
    eprint = "2502.13327",
    archivePrefix = "arXiv",
    primaryClass = "gr-qc",
    doi = "10.1140/epjc/s10052-025-14427-z",
    journal = "Eur. Phys. J. C",
    volume = "85",
    number = "7",
    pages = "724",
    year = "2025"
}

@article{Fabris:2023opv,
    author = "Fabris, J\'ulio C. and Junior, Ednaldo L. B. and Rodrigues, Manuel E.",
    title = "{Generalized models for black-bounce solutions in f(R) gravity}",
    eprint = "2310.00714",
    archivePrefix = "arXiv",
    primaryClass = "gr-qc",
    doi = "10.1140/epjc/s10052-023-12022-8",
    journal = "Eur. Phys. J. C",
    volume = "83",
    number = "10",
    pages = "884",
    year = "2023"
}

@article{Rois:2024qzm,
    author = "R\'ois, Gabriel I. and Junior, Jos\'e Tarciso S. S. and Lobo, Francisco S. N. and Rodrigues, Manuel E.",
    title = "{Novel electrically charged wormhole, black hole, and black bounce exact solutions in hybrid metric-Palatini gravity}",
    eprint = "2412.10324",
    archivePrefix = "arXiv",
    primaryClass = "gr-qc",
    doi = "10.1103/PhysRevD.111.124012",
    journal = "Phys. Rev. D",
    volume = "111",
    number = "12",
    pages = "124012",
    year = "2025"
}

@article{Junior:2023qaq,
    author = "Junior, Jos\'e Tarciso S. S. and Rodrigues, Manuel E.",
    title = "{Coincident $f(\mathbb {Q})$ gravity: black holes, regular black holes, and black bounces}",
    eprint = "2306.04661",
    archivePrefix = "arXiv",
    primaryClass = "gr-qc",
    doi = "10.1140/epjc/s10052-023-11660-2",
    journal = "Eur. Phys. J. C",
    volume = "83",
    number = "6",
    pages = "475",
    year = "2023"
}

@article{Junior:2024cbb,
    author = "Junior, Ednaldo L. B. and Junior, Jos\'e Tarciso S. S. and Lobo, Francisco S. N. and Rodrigues, Manuel E. and Rubiera-Garcia, Diego and da Silva, Lu\'\i{}s F. Dias and Vieira, Henrique A.",
    title = "{Black bounces in Cotton gravity}",
    eprint = "2407.21649",
    archivePrefix = "arXiv",
    primaryClass = "gr-qc",
    doi = "10.1140/epjc/s10052-024-13568-x",
    journal = "Eur. Phys. J. C",
    volume = "84",
    number = "11",
    pages = "1190",
    year = "2024"
}

@article{Junior:2024vrv,
    author = "Junior, Jos\'e Tarciso S. S. and Lobo, Francisco S. N. and Rodrigues, Manuel E.",
    title = "{Black bounces in conformal Killing gravity}",
    eprint = "2405.09702",
    archivePrefix = "arXiv",
    primaryClass = "gr-qc",
    doi = "10.1140/epjc/s10052-024-12922-3",
    journal = "Eur. Phys. J. C",
    volume = "84",
    number = "6",
    pages = "557",
    year = "2024"
}

@article{Nojiri:2017ncd,
    author = "Nojiri, S. and Odintsov, S. D. and Oikonomou, V. K.",
    title = "{Modified Gravity Theories on a Nutshell: Inflation, Bounce and Late-time Evolution}",
    eprint = "1705.11098",
    archivePrefix = "arXiv",
    primaryClass = "gr-qc",
    reportNumber = "PHYS.REPT.-692-(2017)-1-104, Phys.Rept. 692 (2017) 1-104",
    doi = "10.1016/j.physrep.2017.06.001",
    journal = "Phys. Rept.",
    volume = "692",
    pages = "1--104",
    year = "2017"
}

@article{Nojiri:2024qgx,
    author = "Nojiri, Shin'ichi and Odintsov, S. D.",
    title = "{Black holes and their shadows in F(R) gravity}",
    eprint = "2412.13775",
    archivePrefix = "arXiv",
    primaryClass = "gr-qc",
    reportNumber = "KEK-TH-2674, KEK-Cosmo-0367",
    doi = "10.1016/j.dark.2024.101785",
    journal = "Phys. Dark Univ.",
    volume = "47",
    pages = "101785",
    year = "2025"
}

@article{Elizalde:2020icc,
    author = "Elizalde, E. and Nashed, G. G. L. and Nojiri, S. and Odintsov, S. D.",
    title = "{Spherically symmetric black holes with electric and magnetic charge in extended gravity: physical properties, causal structure, and stability analysis in Einstein\textquoteright{}s and Jordan\textquoteright{}s frames}",
    eprint = "2001.11357",
    archivePrefix = "arXiv",
    primaryClass = "gr-qc",
    doi = "10.1140/epjc/s10052-020-7686-3",
    journal = "Eur. Phys. J. C",
    volume = "80",
    number = "2",
    pages = "109",
    year = "2020"
}

@article{Nojiri:2017kex,
    author = "Nojiri, Shin'ichi and Odintsov, S. D.",
    title = "{Regular multihorizon black holes in modified gravity with nonlinear electrodynamics}",
    eprint = "1708.05226",
    archivePrefix = "arXiv",
    primaryClass = "hep-th",
    doi = "10.1103/PhysRevD.96.104008",
    journal = "Phys. Rev. D",
    volume = "96",
    number = "10",
    pages = "104008",
    year = "2017"
}

@article{Nojiri:2014jqa,
    author = "Nojiri, Shin'ichi and Odintsov, Sergei D.",
    title = {{Instabilities and anti-evaporation of Reissner\textendash{}Nordstr\"om black holes in modified $F(R)$ gravity}},
    eprint = "1405.2439",
    archivePrefix = "arXiv",
    primaryClass = "gr-qc",
    doi = "10.1016/j.physletb.2014.06.070",
    journal = "Phys. Lett. B",
    volume = "735",
    pages = "376--382",
    year = "2014"
}

@article{Nojiri:2010wj,
    author = "Nojiri, Shin'ichi and Odintsov, Sergei D.",
    title = "{Unified cosmic history in modified gravity: from F(R) theory to Lorentz non-invariant models}",
    eprint = "1011.0544",
    archivePrefix = "arXiv",
    primaryClass = "gr-qc",
    doi = "10.1016/j.physrep.2011.04.001",
    journal = "Phys. Rept.",
    volume = "505",
    pages = "59--144",
    year = "2011"
}

@article{Tsukamoto:2021caq,
    author = "Tsukamoto, Naoki",
    title = "{Gravitational lensing by two photon spheres in a black-bounce spacetime in strong deflection limits}",
    eprint = "2105.14336",
    archivePrefix = "arXiv",
    primaryClass = "gr-qc",
    doi = "10.1103/PhysRevD.104.064022",
    journal = "Phys. Rev. D",
    volume = "104",
    number = "6",
    pages = "064022",
    year = "2021"
}

@article{Tsukamoto:2020bjm,
    author = "Tsukamoto, Naoki",
    title = "{Gravitational lensing in the Simpson-Visser black-bounce spacetime in a strong deflection limit}",
    eprint = "2011.03932",
    archivePrefix = "arXiv",
    primaryClass = "gr-qc",
    doi = "10.1103/PhysRevD.103.024033",
    journal = "Phys. Rev. D",
    volume = "103",
    number = "2",
    pages = "024033",
    year = "2021"
}

@article{Shtanov:2022xew,
    author = "Shtanov, Yuri",
    title = "{Initial conditions for the scalaron dark matter}",
    eprint = "2207.00267",
    archivePrefix = "arXiv",
    primaryClass = "astro-ph.CO",
    doi = "10.1088/1475-7516/2022/10/079",
    journal = "JCAP",
    volume = "10",
    pages = "079",
    year = "2022"
}

@article{appleby2009curing,
  title={Curing singularities in cosmological evolution of F(R) gravity},
  author={Appleby, Stephen A and Battye, Richard A and Starobinsky, Alexei A},
  journal={arXiv preprint arXiv:0909.1737},
  year={2009}
}

@article{gannouji2012generic,
  title={Generic f(R) theories and classicality of their scalarons},
  author={Gannouji, Rachid},
  journal={Physics Letters B},
  volume={718},
  number={1},
  pages={1--5},
  year={2012},
  publisher={Elsevier}
}

@article{gorbunov2014scalaron,
  title={Scalaron production in contracting astrophysical objects},
  author={Gorbunov, Dmitry S and Tokareva, Irina V},
  journal={arXiv preprint arXiv:1412.3413},
  year={2014}
}

@article{gannouji2012quantum,
  title={Quantum stability bound on the mass of scalaron in generic theories of f(R) gravity},
  author={Gannouji, Rachid},
  journal={arXiv preprint arXiv:1206.3395},
  year={2012}
}

@article{aldabergenov2018beyond,
  title={Beyond Starobinsky inflation},
  author={Aldabergenov, Yermek and others},
  journal={Physical Review D},
  volume={98},
  number={8},
  pages={083511},
  year={2018},
  publisher={APS}
}

@article{Novello2000,
  author    = {M{\'a}rio Novello and V. A. De Lorenci and J. M. Salim and R. Klippert},
  title     = {Geometrical aspects of light propagation in nonlinear electrodynamics},
  journal   = {Physical Review D},
  volume    = {61},
  pages     = {045001},
  year      = {2000},
  doi       = {10.1103/PhysRevD.61.045001}
}

@article{Novello2001,
  author    = {M{\'a}rio Novello and J. M. Salim},
  title     = {Effective electromagnetic geometry},
  journal   = {Physical Review D},
  volume    = {63},
  pages     = {083511},
  year      = {2001},
  doi       = {10.1103/PhysRevD.63.083511}
}

\end{document}